\documentclass[aps,prl,twocolumn,superscriptaddress,longbibliography]{revtex4-2}
\usepackage{graphicx}
\usepackage{amssymb,amsmath}
\usepackage{bm}
\usepackage{dcolumn}
\usepackage{subfigure}
\usepackage{float}
\usepackage[OT1]{fontenc} 
\usepackage{url}
\usepackage{mathrsfs}
\usepackage{slashed}
\usepackage{color}
\usepackage{verbatim}
\usepackage{enumitem}
\usepackage{txfonts}
\usepackage{soul,physics}
\usepackage[driverfallback=dvipdfm]{hyperref}
\hypersetup{pdfpagemode=FullScreen,colorlinks=true,breaklinks,urlcolor=blue,linkcolor=blue,citecolor=blue}
\usepackage{natbib}
\usepackage{environ}

\usepackage{subfigure}
\usepackage{comment}
\usepackage{hyperref}
\usepackage{amssymb}
\usepackage{bm}
\usepackage{graphicx}
\usepackage{amsmath,amssymb}
\usepackage{tikz,fp}
\usepackage{tikz-cd}
\usetikzlibrary{arrows}
\usetikzlibrary{intersections}
\usetikzlibrary{shapes.geometric}
\usetikzlibrary{decorations.pathmorphing, patterns,shapes,fixedpointarithmetic}
\usetikzlibrary{decorations.markings}

\def\be{\begin{equation}}
\def\ee{\end{equation}}
\def\bea{\begin{eqnarray}}
\def\eea{\end{eqnarray}}

\usepackage{tikz,fp}
\usepackage{tikz-cd}
\usetikzlibrary{arrows}
\usetikzlibrary{intersections}
\usetikzlibrary{shapes.geometric}
\usetikzlibrary{decorations.pathmorphing, patterns,shapes,fixedpointarithmetic}
\usetikzlibrary{decorations.markings}

\pgfdeclarepatternformonly{south west lines}{\pgfqpoint{-0pt}{-0pt}}{\pgfqpoint{3pt}{3pt}}{\pgfqpoint{3pt}{3pt}}{
  \pgfsetlinewidth{0.4pt}
  \pgfpathmoveto{\pgfqpoint{0pt}{0pt}}
  \pgfpathlineto{\pgfqpoint{3pt}{3pt}}
  \pgfpathmoveto{\pgfqpoint{2.8pt}{-.2pt}}
  \pgfpathlineto{\pgfqpoint{3.2pt}{.2pt}}
  \pgfpathmoveto{\pgfqpoint{-.2pt}{2.8pt}}
  \pgfpathlineto{\pgfqpoint{.2pt}{3.2pt}}
  \pgfusepath{stroke}}

\tikzset{
  mid arrow/.style={postaction={decorate,decoration={
        markings,
        mark=at position .575 with {\arrow{stealth}}
  }}},
  near arrow/.style={postaction={decorate,decoration={
        markings,
        mark=at position .275 with {\arrow{stealth}}
  }}},
  far arrow/.style={postaction={decorate,decoration={
        markings,
        mark=at position .800 with {\arrow{stealth}}
  }}},
  snake arrow/.style={fixed point arithmetic, decorate, decoration={snake,amplitude=2pt, segment length=11pt},postaction={decoration={markings,mark=at position 0.625 with {\arrow{stealth}}},decorate}},
}
\tikzset{
  baseline = -0.5ex,
  wavy/.style = {
    thick,
    decorate,
    decoration={snake,amplitude=2pt,segment length=5pt}},
  sdot/.style = {
    circle,
    draw=none,
    fill=black,
    minimum size=2.5pt,
    inner sep=0pt},
  bdot/.style = {
    circle,
    draw=none,
    fill=black,
    minimum size=4pt,
    inner sep=0pt},
  svertex/.style = {
    circle,
    draw=black,
    thick,
    fill=lightgray,
    minimum size=8pt,
    inner sep=1pt},
  bvertex/.style = {
    circle,
    draw=black,
    thick,
    fill=lightgray,
    minimum size=24pt},
  bvertexsmall/.style = {
    circle,
    draw=black,
    thick,
    fill=lightgray,
    minimum size=7pt},
  bvertexnormal/.style = {
    circle,
    draw=black,
    thick,
    fill=lightgray,
    minimum size=14pt},
  dvertex/.style = {
    circle,
    draw=black,
    thick,
    fill=gray,
    minimum size=25pt}}

\begin{document}
  
\hfill{\footnotesize USTC-ICTS/PCFT-24-12}

\title{Dissecting Quantum Many-body Chaos in the Krylov Space}
  
  \author{Liangyu Chen}
\affiliation{Kavli Institute for Theoretical Sciences, University of Chinese Academy of Sciences, Beijing 100190, China}
  
  \author{Baoyuan Mu}
\affiliation{Kavli Institute for Theoretical Sciences, University of Chinese Academy of Sciences, Beijing 100190, China}

  \author{Huajia Wang}
  \thanks{wanghuajia@ucas.ac.cn}
\affiliation{Kavli Institute for Theoretical Sciences, University of Chinese Academy of Sciences, Beijing 100190, China}
\affiliation{Peng Huanwu Center for Fundamental Theory, Hefei, Anhui 230026, China.}

  \author{Pengfei Zhang}
  \thanks{PengfeiZhang.physics@gmail.com}
  \affiliation{Department of Physics, Fudan University, Shanghai 200438, China}
  \affiliation{Shanghai Qi Zhi Institute, AI Tower, Xuhui District, Shanghai 200232, China}
  \affiliation{Hefei National Laboratory, Hefei, 230088, China}
  \date{\today}

  \begin{abstract}
  The growth of simple operators is essential for the emergence of chaotic dynamics and quantum thermalization. Recent studies have proposed different measures, including the out-of-time-order correlator and Krylov complexity. It is established that the out-of-time-order correlator serves as the signature of quantum many-body chaos, while the Krylov complexity provides its upper bound. However, there exist non-chaotic systems in which Krylov complexity grows exponentially, indicating that the Krylov complexity itself is not a witness of many-body chaos. In this letter, we introduce the missing ingredient, named as the Krylov metric $K_{mn}$, which probes the size of the Krylov basis. We propose that the universal criteria for fast scramblers include (i) the exponential growth of Krylov complexity, (ii) the diagonal elements $K_{nn}\sim n^h$ with $h\in(0,1]$, and (iii) the negligibility of off-diagonal elements $K_{mn}$ with $m\neq n$. We further show that $h=\varkappa / 2\alpha$ is a ratio between the quantum Lyapunov exponent $\varkappa$ and the Krylov exponent $\alpha$. This proposal is supported by both generic arguments and explicit examples, including solvable SYK models, Luttinger Liquids, and many-body localized systems. Our results provide a refined understanding of how chaotic dynamics emerge from the Krylov space perspective.
  \end{abstract}
  
  \maketitle

  \emph{ \color{blue}Introduction.--} 
  Understanding how chaotic dynamics emerge and drive systems toward thermal equilibrium is of vital importance in the study of quantum dynamics. It requires encoding all local initial conditions into the entire system after sufficiently long unitary evolutions, a phenomenon known as information scrambling \cite{Hayden:2007cs,Sekino:2008he}. Inspired by gravity calculations \cite{DRAY1985173,THOOFT1990138,Kiem:1995iy,tHooft:1996rdg}, the out-of-time-order correlator (OTOC) is introduced as a quantitative measure of quantum many-body chaos \cite{1969JETP...28.1200L,Shenker:2013pqa,Shenker:2014cwa,kitaev2014talk,Roberts:2014isa}, which provides broad implications in condensed matter physics, quantum information, and high-energy physics. It probes the average operator size, which is a measure of operator complexity in the local basis \cite{Roberts:2014isa,Roberts:2018mnp,Qi:2018bje}. Of particular interest are chaotic systems with large local Hilbert space dimensions, wherein the OTOC exhibits exponential deviation behavior characterized by quantum Lyapunov exponent $\varkappa$ \cite{Shenker:2013pqa,Shenker:2014cwa,kitaev2014talk,Roberts:2014isa,Maldacena:2015waa}. Examples include the Sachdev-Ye-Kitaev (SYK) model \cite{2015escq.progE...2K,PhysRevLett.70.3339,Maldacena:2016hyu,Kitaev:2017awl,Chowdhury:2021qpy}, Brownian circuits \cite{2024PRXQ....5a0201X,Chen:2018bqy,Zhou:2018snw}, and black holes \cite{Shenker:2013pqa, Shenker:2014cwa, Maldacena:2017axo, Roberts:2014isa}, often referred to as fast scramblers \cite{Sekino:2008he}.

  In addition to the OTOC or operator size, various measures of operator complexity have been proposed by selecting different bases \cite{Roberts:2016hpo,Jefferson:2017sdb,Yang:2017nfn,Chapman:2017rqy,Khan:2018rzm,Yang:2018nda,Lucas:2018wsc,Balasubramanian:2019wgd,Balasubramanian:2021mxo}. The Krylov basis stands out because it provides a convenient and intrinsically defined operator basis generated by the Heisenberg evolution \cite{Parker:2018yvk,2022arXiv221214429A,Balasubramanian:2022tpr,2023PhRvR...5c3085L,Barbon:2019wsy,Dymarsky:2019elm,Barbon:2019tuq,Magan:2020iac,Jian:2020qpp,Rabinovici:2020ryf,Chen:2019klo,2021PhRvE.104c4112N,Caputa:2021ori,Patramanis:2021lkx,Bhattacharjee:2022vlt,Caputa:2021sib,Dymarsky:2021bjq,Avdoshkin:2019trj,Rabinovici:2023yex,Liu:2022god,Bhattacharya:2022gbz,Bhattacharjee:2022lzy,Bhattacharjee:2023uwx,2023arXiv230307343L,2023arXiv231217416T,2023arXiv230502356Z}. In this basis, the operator dynamics are mapped to the evolution of a wavepacket on a half-infinite chain with nearest-neighbor hopping determined by the Lanczos coefficients $b_n$. The Krylov complexity $\mathcal{K}(t)$ is then defined as the center-of-mass position at time $t$. In Ref. \cite{Parker:2018yvk}, it is conjectured that Lanczos coefficients approach a linear form $b_n=\alpha n+\gamma$ at large $n$ for chaotic systems, which gives $\mathcal{K}(t)\propto e^{2\alpha t}$. Furthermore, the Krylov exponent provide an upper bound to the quantum Lyapunov exponent as $\varkappa\leq 2\alpha$ \cite{Parker:2018yvk,Avdoshkin:2019trj}, saturated in the large-$q$ SYK model. 
  However, there are instances that exhibit significant discrepancies between the Krylov complexity and the actual chaotic/integrable nature of the system captured by OTOCs. The extreme examples include the Krylov complexity of local operators in a free 2D conformal field theory (CFT) \cite{Dymarsky:2021bjq}, which grows exponentially despite the system being non-interacting.

    \begin{figure}[tb]
    \centering
    \includegraphics[width=0.92\linewidth]{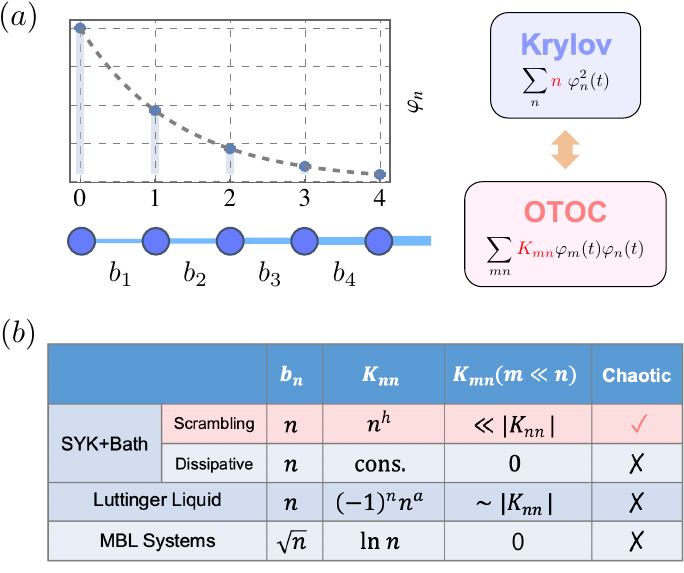}
    \caption{ (a). In the Krylov basis, the Heisenberg evolution is mapped to a tight-binding model with nearest neighbor hopping. The Krylov complexity measures the position of the wavepacket, while the equal-time OTOC measures a non-local operator $K_{mn}$. Here, we assumed the operator $\hat{\mathcal{O}}$ is Hermitian. (b). We summarize the results from the SYK models, Luttinger Liquids, and many-body localized systems, which support our criteria.   }
    \label{fig:summary}
  \end{figure}

  This naturally leads to the following question: What is the criterion for diagnosing quantum many-body chaos in the Krylov space? In this letter, we attempt to elucidate this issue by explicitly representing the OTOC as a non-local observable in the Krylov basis, denoted as a Krylov metric $K_{mn}$, which measures the size of Krylov basis. The difference between the Krylov complexity and OTOC is then reflected in the distinction between the position operator $n\delta_{mn}$ and $K_{mn}$ for large $m$ and $n$. We provide the hypothesis for fast scramblers: (i) the Krylov complexity grows exponentially, (ii) the diagonal elements $K_{nn}$ show power-law increase $n^{h} = n^{\varkappa /2\alpha}$, and (iii) the off-diagonal elements of $K_{mn}$ are negligible. This proposal is supported by examples including the SYK models \cite{2015escq.progE...2K,PhysRevLett.70.3339,Maldacena:2016hyu,Kitaev:2017awl,Chowdhury:2021qpy}, Luttinger Liquids \cite{10.1093/acprof:oso/9780198525004.001.0001}, and many-body localized (MBL) systems \cite{RevModPhys.91.021001,2018CRPhy..19..498A,2015ARCMP...6..383A,2015ARCMP...6...15N,PhysRevLett.111.127201,PhysRevB.90.174202,PhysRevLett.110.067204}, as summarized in FIG. \ref{fig:summary}.

  \emph{ \color{blue}OTOC in Krylov Space.--} The Krylov basis is defined with respect to a simple operator $\hat{\mathcal{O}}$ and Hamiltonian $\hat{H}$. Utilizing the operator-state mapping, we express $\hat{\mathcal{O}}$ as a state $|\mathcal{O}\rangle$ in the doubled Hilbert space. The Heisenberg equation takes the form $d|\mathcal{O}\rangle/dt=\mathcal{L}~|\mathcal{O}\rangle$, where $\mathcal{L}:= i[H,\cdot]$ denotes the Liouvillian superoperator \footnote{Notice that our convention for the Liouvillian superoperator differs from that in \cite{Parker:2018yvk} by a factor of $i$.}. The time evolution couples $|\mathcal{O}\rangle$ to a set of operators $\{\mathcal{L}^n~|\mathcal{O}\rangle\}$ with $n\in\{0,1,2,...\}$. Applying the Gram-Schmidt procedure with respect to the inner product $\langle \mathcal{O}_1|\mathcal{O}_2\rangle=\langle \mathcal{O}_1^\dagger \mathcal{O}_2\rangle =\text{tr}[\mathcal{O}_1^\dagger \mathcal{O}_2]/\text{tr}[1]$, we obtain the recursive construction for the orthonormal Krylov basis ($n\geq 2$)
  \begin{equation}\label{eqn:Krylov construction}
  \begin{aligned}
  |A_n\rangle:=&\mathcal{L} |\mathcal{O}_{n-1}\rangle+b_{n-1}|\mathcal{O}_{n-2}\rangle,\\
  |\mathcal{O}_n\rangle:=& b_n^{-1}|A_n\rangle,\ \ \ \ \ \ \ \ \ \ \ \ b_n:={\langle A_n|A_n\rangle}^{1/2}.
  \end{aligned}
  \end{equation}
  Here, initial conditions are $|\mathcal{O}_0\rangle=|\mathcal{O}\rangle$ and $|\mathcal{O}_1\rangle=b_1^{-1}\mathcal{L}~|\mathcal{O}_0\rangle $. We have assumed that $|\mathcal{O}\rangle$ is normalized and $b_1$ is the normalization factor for $|\mathcal{O}_1\rangle$. The set of positive numbers $\{b_n\}$ are called the Lanczos coefficients \cite{Lanczos:1950zz}, which only depends on two-point functions $G(t)=\langle \mathcal{O}|\mathcal{O}(t)\rangle$. Eq. \eqref{eqn:Krylov construction} indicates that the Liouvillian superoperator is nearest-neighbour in the Krylov basis. Introducing $\hat{\mathcal{O}}(t) =\sum_n\varphi_n(t)~\hat{\mathcal{O}}_n$, the evolution of operator wavefunction $\varphi_n(t)$ is mapped to a tight binding model with
  ${d\varphi_n(t)}/{dt}=b_n \varphi_{n-1}(t)-b_{n+1}\varphi_{n+1}(t).$ The Krylov complexity is defined as the expectation of the position operator $\mathcal{K}(t):=\sum_n n~|\varphi_n(t)|^2$, which measures how fast the wavepacket spreads towards larger $n$. Ref. \cite{Parker:2018yvk} proposed the hypothesis that $b_n$ approaches a linear function $b_n=\alpha n+\gamma$ at large $n$ for chaotic systems in the thermodynamic limit, which results in $\mathcal{K}(t) \sim e^{2\alpha t}$.
  
  We are interested in the relation between Krylov complexity and the OTOC. Our main focus is on its connected part:
  \begin{equation}
  F(t_1,t_2)=G(t_{12})\mp\langle\hat{\mathcal{O}}(t_2)^\dagger\hat{\mathcal{O}}'(0)^\dagger\hat{\mathcal{O}}(t_1)\hat{\mathcal{O}}'(0)\rangle,
  \end{equation}
  where $t_{12}=t_1-t_2$ and the positive sign is chosen when both $\hat{\mathcal{O}}$ and $\hat{\mathcal{O}}'$ are fermionic, and we assume the normalization $\langle \hat{\mathcal{O}}'^2\rangle=1$. In chaotic systems with large local Hilbert space dimensions, the OTOC exhibits exponential deviation behavior $F(t_1,t_2)\sim f(t_{12})e^{\varkappa T_{12}}$ with $T_{12}=(t_1+t_2)/2$ until the scrambling time. It is known that the quantum Lyapunov exponent $\varkappa$ is bounded by the Krylov exponent as $\varkappa \leq 2\alpha$. We rewrite the OTOC by expressing $\mathcal{O}(t_i)$ using the operator wavefunction:
  \begin{equation}\label{eqn:Fexpansion}
  \begin{aligned}
  &F(t_1,t_2)=\sum_{mn}K_{mn}~\varphi_m(t_1)\varphi_n(t_2)^*,\\
  &K_{mn}=\delta_{mn}\mp \langle\hat{\mathcal{O}}_n^\dagger\hat{\mathcal{O}}'(0)^\dagger\hat{\mathcal{O}}_m\hat{\mathcal{O}}'(0)\rangle.
  \end{aligned}
  \end{equation}
  We refer to $K_{mn}$ as the Krylov metric. Notice that although we focused on the infinite temperature limit in previous discussions, the effects of finite temperature can be incorporated naturally by replacing $\hat{\mathcal{O}}$ with $\rho^{\frac{1}{4}}\hat{\mathcal{O}}\rho^{\frac{1}{4}}$ \cite{Qi:2018bje}. Here, $\rho=e^{-\beta \hat{H}}/\text{tr}[e^{-\beta \hat{H}}]$ is the thermal density matrix. Under this replacement, the two-point function matches the Wightman Green's function  $G(t)=\langle \rho^{\frac{1}{2}}\hat{\mathcal{O}}(t)\rho^{\frac{1}{2}}\hat{\mathcal{O}}\rangle$, and the OTOC exhibits equal imaginary-time separations. 

  \emph{ \color{blue}Generic Analysis.--} Eq. \eqref{eqn:Fexpansion} reveals that the distinction between the OTOC and the Krylov complexity lies in how they measure the Krylov space: The spread of the operator wavefunction depends solely on the Lanczos coefficients, which exhibit an exponential behavior for $b_n\sim \alpha n$. However, this behavior does not necessarily imply quantum many-body chaos for a generic $K_{mn}$. The manifestation of many-body chaos requires the ability of $K_{mn}$ to measure the spreading in the Krylov space, indicating $K_{nn}$ as an increasing function of $n$. Physically, as the OTOC measures the growth of operator size in time, the Krylov metric measures the size growth in the Krylov index $n$. For illustration, let us consider systems that consist of Majorana fermions $\hat{\chi}_j$ with $\{\hat{\chi}_j,\hat{\chi}_k\}=2\delta_{jk}$. Choosing $\hat{\mathcal{O}}'=\hat{\chi}_j$, Eq. \eqref{eqn:Fexpansion} becomes \cite{Roberts:2018mnp,Qi:2018bje}
  \begin{equation}\label{eqn:size}
  \overline{K_{nn}}=\frac{1}{2N}\sum_j\left<\left|[\hat{\mathcal{O}}_n,\hat{\chi}_j]\right|^2\right>=\frac{1}{2N}\text{Size}[\hat{\mathcal{O}}_n].
  \end{equation}
  Here, we averaged $K_{nn}$ over different $j$, which is unnecessary for SYK-like models with permutation symmetry between different Majorana modes. Similarly, off-diagonal components $K_{mn}$ measure the interference between $\hat{\mathcal{O}}_m$ and $\hat{\mathcal{O}}_n$ weighted by the operator size. It is also straightforward to generalize this relation to spin systems \cite{2024PhRvL.132f0201L}. 

  To further motivate the criteria for chaotic systems with exponentially growing OTOC, let's employ a naive scaling argument. Firstly, since $\varkappa$ is bounded by $\alpha$, a non-vanishing $\varkappa$ requires exponential growth of $\mathcal{K}(t)$. In this scenario, the Krylov complexity suggests the identification $n\sim e^{2\alpha t}$. Applying this relation to the OTOC, we find $F(t,t)\sim e^{\varkappa t}\sim n^{h}\sim K_{nn}$. Additionally, the validity of this argument necessitates the Krylov metric to be approximately diagonal, i.e., $|K_{mn}|\ll |K_{mm}|$ for $m\gg n$. Otherwise, the double summation in \eqref{eqn:Fexpansion} leads to additional enhancement due to off-diagonal coherence. Physically, this occurs because operators with different Krylov indices have different typical sizes, thus their overlap is significantly smaller than diagonal elements. This leads to our universal fast scrambler hypothesis
  \begin{enumerate}
  \item The Lanczos coefficients approach a linear function $b_n=\alpha n$, allowing the Krylov complexity to grow exponentially.

  \item The diagonal elements $K_{nn}$ are proportional to $n^h$, allowing it to measure the spreading in the Krylov space.

  \item The off-diagonal elements $K_{mn}$ $(m\neq n)$ are negligible, allowing the scaling argument to hold.

  \end{enumerate}

 \emph{ \color{blue}Example 1: The SYK Model.--} To support our proposal, we first study the Krylov metric in the SYK model \cite{2015escq.progE...2K,PhysRevLett.70.3339,Maldacena:2016hyu,Kitaev:2017awl,Chowdhury:2021qpy}. In order to achieve a tunable quantum Lyapunov exponent, we couple system fermions, denoted as $\hat{\chi}_j$ ($j=1,2,...,N$), to a series of bath fermions, denoted as $\hat{\psi}_a$ ($a=1,2,...,M$) with $M\gg N$ \cite{Chen:2017dbb,Zhang:2023xrr}. The Hamiltonian reads 
 \begin{equation}
 \begin{aligned}
 H=&\sum_{i<j<k<l}J_{ijkl}\hat{\chi}_i\hat{\chi}_j\hat{\chi}_k\hat{\chi}_l+\sum_{a<b<c<d}J_{abcd}'\hat{\psi}_a\hat{\psi}_b\hat{\psi}_c\hat{\psi}_d\\
 &+\sum_{i<j}\sum_{a<b}u_{ijab}\hat{\chi}_i\hat{\chi}_j\hat{\psi}_a\hat{\psi}_b.
 \end{aligned}
 \end{equation}
 Here, the coupling strengths $J_{ijkl}$, $J_{abcd}'$, and $u_{ijab}$ are independent Gaussian varibles with zero means and 
 \begin{equation}
 \overline{J_{ijkl}^2}=\frac{6J^2}{N^3},\ \ \ \ \ \  \overline{J_{abcd}'^2}=\frac{6J^2}{M^3}\ \ \ \ \ \ \overline{u_{ijab}^2}= \frac{2u^2}{N M^2}.
 \end{equation}

 The model has been analyzed using the large-$N$ expansion in the low-energy limit with $\beta J \gg 1$ \cite{Chen:2017dbb}. The (normalized) two-point function for $\hat{\mathcal{O}}\sim \hat{\chi}_1$ is given by $G(t)={(\cosh(\alpha t))^{-2\Delta}}$, with $\alpha=\pi/\beta$ and $\Delta=1/4$. Models with a generic $\Delta$ can be constructed following a similar strategy. Importantly, the two-point function remains independent of the dimensionless system-bath coupling $u/J$. Therefore, the Lanczos coefficients and the operator wavefunction match those of the traditional SYK model, which read \cite{Parker:2018yvk,Caputa:2021sib}
\begin{equation}\label{phi}
\begin{aligned}
b_n&=\alpha\sqrt{n(n+2\Delta-1)},\ \ \ \ \ \ \ \ \ 
\varphi_n(t) = D_n\frac{\tanh{(\alpha t)}^n}{\cosh{(\alpha t)}^{2\Delta}} ,
\end{aligned}
\end{equation}
  where $D_n=\sqrt{\frac{\Gamma(2\Delta+n)}{\Gamma(n+1)\Gamma(2\Delta)} }$. This identifies $\alpha$ as the Krylov exponent. The OTOC with $\hat{\mathcal{O}}'\sim\hat{\chi}_2$ can be computed by summing up ladder diagrams \cite{Maldacena:2016hyu}. The result reads
  \begin{equation}
 F(t_1,t_2)=f(t_{12})e^{\varkappa T_{12}}=C_0 \frac{e^{2\alpha h T_{12}}}{\cosh(\alpha t_{12})^{2\Delta+h}},
 \end{equation}
 where the Lyapunov exponent $\varkappa=2h\alpha$ exhibits explicit $u/J$ dependence through $h=\left( 1-\frac{\sqrt{k^4+4k^2}-k^2}{2}\right)$ with $k=u^2/J^2$ \cite{Chen:2017dbb}. Particularly, in the limit as $k\rightarrow \infty$, the Lyapunov exponent $\varkappa \rightarrow 0$, indicating the system transitions into a non-chaotic dissipative phase \cite{Zhang:2023xrr}.
 
\begin{figure}[t]
    \centering
    \includegraphics[width=0.99\linewidth]{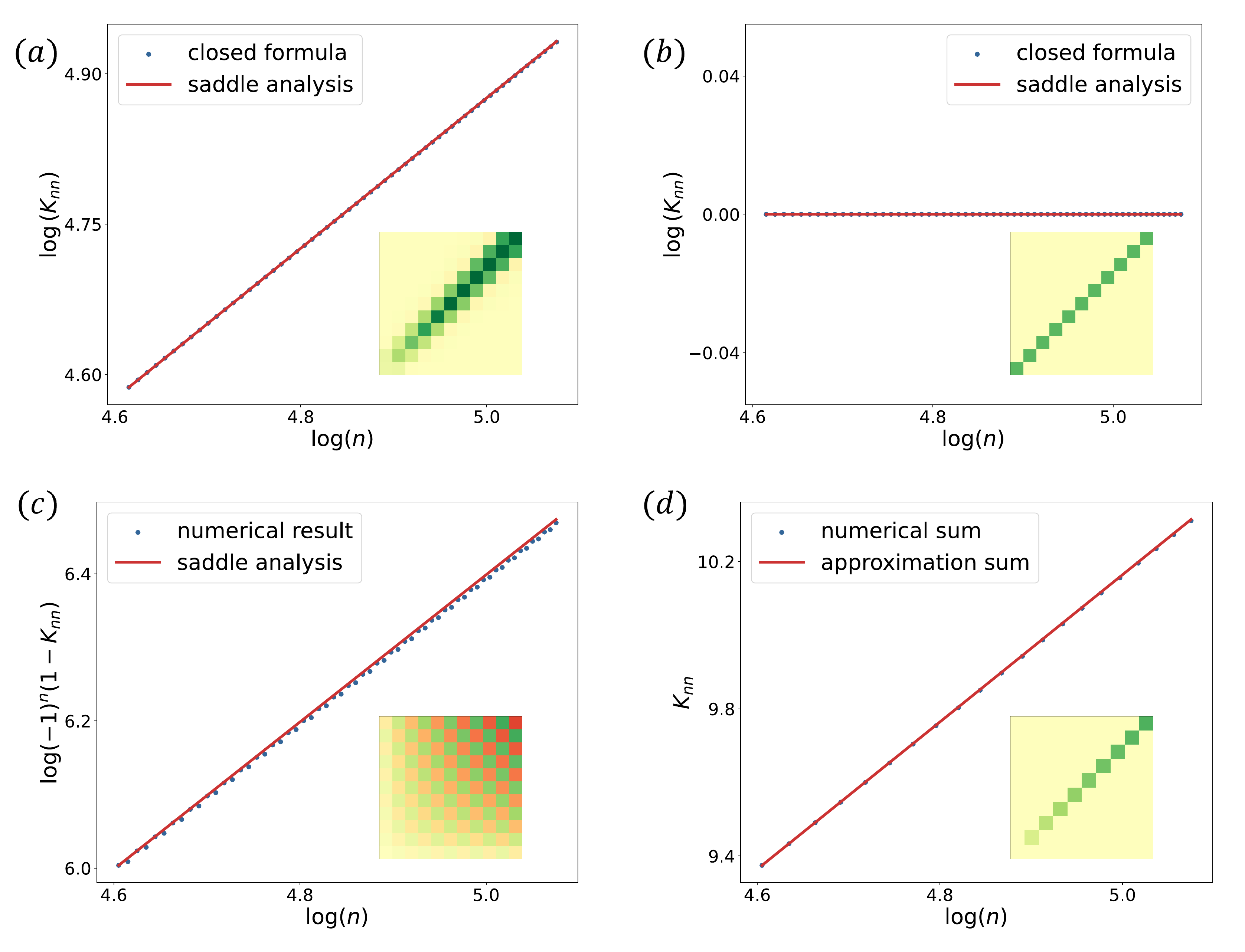}
    \caption{We present the plot of the Krylov metric for (a) the SYK model with $h=3/4$, (b) the SYK model with $h=0$, (c) the Luttinger liquids with $\Delta=1$, and (d) MBL systems with $\xi=1$. In each sub-figure, we also show a matrix plot of $K_{mn}$ as an inset, where yellow, red and green represent zero, negative and positive numbers, respectively. For Luttinger liquids, the matrix plot only contains $K_{mn}$ with even indices so that they are real.
     }
    \label{fig:KmnPlot}
 \end{figure}

 We explore the origin of the discrepancy between $\varkappa$ and $2\alpha$ by computing the Krylov metric $K_{mn}$. Utilizing the analytical knowledge of both the OTOC and the operator wavefunction, we introduce the auxiliary variable $y_i=\tanh(\alpha t_i)$:
\begin{equation}\label{eqn:tobeexpand}
\left[\prod_{i}\cosh{(\alpha t_i)}^{2\Delta}\right]F(t_1,t_2) =C_0\frac{(1+y_1)^h(1+y_2)^h}{(1-y_1y_2)^{2\Delta+h}}.
\end{equation}
Accorcding to \eqref{eqn:Fexpansion}, its Taylor expansion in $y_1$ and $y_2$ should be matched with $\sum_{mn}D_mD_nK_{mn}y_1^my_2^n$. The result can be computed in closed-form. We leave the complete expression in the supplementary material \cite{SM}, where we confirmed that $K_{mn}$ is dominated by the diagonal element
\begin{equation} 
\begin{aligned}
K_{nn}\propto\frac{ \Gamma (h+n+2 \Delta ) \, _3F_2(-h,-h,-n;1,-h-n-2 \Delta
   +1;1)}{\Gamma (h+2 \Delta ) \Gamma (n+2 \Delta )}.
\end{aligned}
\end{equation}
Here, $\, _3F_2$ represents the generalized hypergeometric function. When we expand the result in the limit of $n \rightarrow \infty$, we derive the following asymptotic behaviors (also see FIG. \ref{fig:KmnPlot} (a-b)): (i) the diagonal elements satisfy: $K_{nn} \propto n^h = n^{\varkappa/(2\alpha)}$, notice this produces a constant $K_{nn}$ in the dissipative limit $u/J \rightarrow 0$, where the quantum Lyapunov exponent vanishes, while the Krylov exponent is $\alpha=2\pi/\beta$;  (ii) for generic $0<h<1$, the off-diagonal elements scale as: $K_{n+m,n-m} \propto K_{nn} m^{-2h-1}$ for $n\gg m\gg 1$, i.e. they exhibit power-law decay along the orthogonal off-diagonal direction; (iii) in the limit of either the dissipative or maximally chaotic phase $h\to \lbrace 0,1\rbrace$, the Krylov metric $K_{mn}$ approaches being exactly diagonal. These behaviors can also be explicitly derived through a saddle-point analysis \cite{SM}.

\emph{ \color{blue}Example 2: Luttinger Liquids.--} We now turn to the investigation of free CFTs, as an extreme example of non-chaotic systems. We consider Luttinger liquids, which describes a large class of gapless quantum matters in 1+1D \cite{10.1093/acprof:oso/9780198525004.001.0001}. The Hamiltonian reads
\begin{equation}
    \begin{aligned}
         H &= \frac{u}{2 \pi} \int d x \left[ \frac{1}{K}(\nabla \hat \phi(x))^2 + K (\pi \hat \Pi(x))^2\right].
    \end{aligned}
 \end{equation}
where $K$ is the Luttinger parameter and $u$ is the sound velocity. $\hat \phi(x)$ is a scalar field with conjugate momentum $\hat \Pi(y)$, which satisfies the canonical commutation relation $[\hat \phi(x),\hat \Pi(y)]=i\delta(x-y)$. We focus on the vertex operator $\hat{\mathcal{O}}\sim :e^{-i n_0 \hat{\phi}(x)}:$, where $:\ :$ denotes normal ordering. The two-point function is determined by the conformal symmetry $G(t)={(\cosh(\alpha t))^{-2\Delta}}$ with scaling dimension $\Delta = {K n_0^{2}}/{4}$. Since this two-point function takes the same form as in the SYK model, the Lanczos coefficients and the operator wavefunction are still given by \eqref{phi}. We then proceed to compute the OTOC. Taking $\hat{\mathcal{O}'}\sim :e^{i n_0 \hat{\phi}(x)}:$, the result of OTOC reads:
\begin{equation}\label{eqn:LL}
  F(t_1,t_2)=G(t_{12})\left(1- \left[ \frac{\cosh\left(\alpha t_{12}\right) -i \sinh\left(2\alpha T_{12}\right)}{\cosh\left(\alpha t_{12}\right) +i \sinh\left(2\alpha T_{12}\right) } \right]^{ 2 \Delta }\right).
  \end{equation}

The Krylov metric $K_{mn}$ is computed using a strategy similar to that of Eq. \eqref{eqn:tobeexpand}. Since the conotribution from the first term in \eqref{eqn:LL} is $\delta_{mn}$, we only focus on the second term. Unfortunately, we are unable to compute its expansion in closed form. Instead, we represent it as a contour integral:
\begin{equation}
\label{eq:kmn_contour}
K_{mn} = \frac{D^{-1}_{m}D^{-1}_{n}}{(2\pi i)^2}\oint_C \frac{dy_1}{y_1^{m+1}} \oint_C \frac{dy_2}{y_2^{n+1}} \Big[\prod_i\cosh{(\alpha t_i)}^{2\Delta}\Big]F(t_1,t_2),
\end{equation}
where the integrand should be viewed as functions of $y_{1,2}$. The integral contour $C$ is along the unit circle, which encloses the origin counterclockwise. We can extract the asymptotic form of $K_{mn}$ for $n,m \gg 1$ by evaluate the contour integral (\ref{eq:kmn_contour}) using the saddle-point approximation. 
 We find that:
 \begin{equation}
 K_{mn} \sim (-1)^{\frac{m+n}{2}} (mn)^{\Delta-1/2}
 \end{equation}
 This result reveals two important observations: (i) The diagonal components $K_{nn}$ exhibit alternating signs when $n$ is changed, leading to a cancellation effect among different $n$ values. (ii) The off-diagonal components $K_{mn}$ are comparable to the diagonal components, see FIG. \ref{fig:KmnPlot} (c). As a result, the summation in \eqref{eqn:Fexpansion} experiences significant cancellation, distinguishing Luttinger liquids from the SYK model. 

 \emph{ \color{blue}Example 3: MBL systems.--} Finally, we consider cases that lie in between chaotic systems and non-interacting systems. A celebrated example is systems exhibiting many-body localization \cite{RevModPhys.91.021001,2018CRPhy..19..498A,2015ARCMP...6..383A,2015ARCMP...6...15N,PhysRevLett.111.127201,PhysRevB.90.174202,PhysRevLett.110.067204}. We examine the effective Hamiltonian \cite{PhysRevLett.111.127201,PhysRevB.90.174202,PhysRevLett.110.067204}
 \begin{equation}\label{eq:MBLeff}
 H=\frac{1}{2} \sum_{i\neq j} J_{ij} \hat{\sigma}^i_z \hat{\sigma}^j_z + \sum_i h_i \hat{\sigma}^i_z+...
 \end{equation}  
 This model is believed to capture the essential features of MBL systems. In particular, it commutes with an extensive number of mutually commuting operators $\sigma^i_z$, known as local integrals of motion (LIOM). $...$ denotes possible higher-order terms, which describes multi-body interactions between LIOMs. For simplicity, we neglect their contributions to correlation functions.  We model random couplings $J_{ij}$ and magnetic field $h_i$ as independent Gaussian varibles with zero means and 
 \begin{equation}
\overline{J_{ij}^2} = J^2 e^{-\frac{|i-j|}{\xi}}, \ \ \ \ \ \ \overline{ h_i^2} = h^2.
 \end{equation}  
 Here, $\xi$ is known as the localization length, which characterize the interaction range between LIOMs. We expect the physical results do not reply on details of the distribution function.
 
 We choose $\hat{\mathcal{O}}=\hat{\sigma}_x^0$ that flips LIOMs. In localized systems, the violation of thermalization renders finite temperature ensembles meaningless. Therefore, all calculations are performed at infinite temperature. The Lanzcos coeffcients are then fixed by the auto-correlation function, which reads
$G(t)=\overline{\prod_{j\neq 0} \cos{(2J_{0j}t)} \cos{(2h_0 t)}}=e^{-\frac{\gamma^2 t^2}{2}}$.
Here, we averaged over the random couplings and introduced $\gamma^2=4(J^2\sum_{j\neq 0}e^{-|j|/\xi}+h^2)$. The result shows that the auto-correlation function decays as a Gaussian function, of which the Krylov basis wave-function and the Lanzcos coefficients are known as \cite{Caputa:2021sib}
\begin{equation}\label{eq:gaussian_krylov}
b_n = \gamma \sqrt{n}\ \ \ \ \ \ \ \ \ \varphi_n(t) = \frac{\gamma^n t^n}{\sqrt{n!}} e^{-\frac{\gamma^2 t^2}{2}},
\end{equation}
and the Krlov complexity grows quadratically $\mathcal{K}(t)=\gamma^2t^2$. We proceed to investigate the behavior of Krylov metric. Choosing $\hat{\mathcal{O}}=\hat{\sigma}_x^m$ and performing an average over site $m$, we find 
\begin{equation}
\begin{aligned}
\overline{F(t_1,t_2)} =& e^{-\frac{\gamma^2t^2_{12}}{2} } - {N}^{-1}{e^{-2{\gamma^2}T_{12}^2}}-\sum_{m\neq 0} N^{-1}e^{-8 J^2 e^{-\frac{|m|}{\xi}} t_1 t_2-\frac{\gamma^2 t_{12}^2}{2}}.
\end{aligned}
\end{equation}
Here, $N$ denotes total number of sites. Using \eqref{eqn:Fexpansion}, we find the Krylov metric takes a purely diagonal form, see Figure (\ref{fig:KmnPlot}): 
\begin{equation}
K_{mn} = K_{nn}\delta_{mn},\;\;K_{nn}\approx \frac{2 \xi}{N}\ln \left(\frac{8J^2n}{\gamma^2}\right).
\end{equation}
This $\ln n$ behavior is a signature of logarithmic lightcones in the MBL system \cite{Huang:2016knw,Fan:2016ean,2017PhRvB..95f0201S,2017PhRvB..95e4201H,2016arXiv160802765C,2017AnP...52900332C}. Although the Krylov complexity grows quadratically, the operator size in MBL systems only increases as $\ln t$. The scaling analysis then suggests $K_{nn}\sim \log n$, supported in FIG. \ref{fig:KmnPlot} (d). This example demonstrates the usefulness of the Krylov metric beyond identifying fast scramblers, indicating its broad application in characterizing quantum dynamics.

\emph{ \color{blue} Size-resolved Metric--} Given the general interpretation of the OTOC in terms of the operator size growth, more refined information of the Krylov metric can be revealed by resolving it using the operator size distribution. In lattice models where operator sizes can be explicitly defined, we can perform the following decomposition: 
\begin{equation}
K_{mn} = \sum_{\ell} K_{mn}(\ell),\;\;K_{mn}(\ell) = \ell \langle \mathcal{O}_m| \hat{P}(\ell)|\mathcal{O}_n\rangle   
\end{equation}
where $\hat{P}(\ell)$ is the projector into the operator Hilbert space sector of fixed size $\ell$. The decomposition $K_{mn}(\ell)$ can be retrieved by first computing and then expanding the operator-size generating functions. The details of these computations can be found in the supplementary material \cite{SM}.
For the SYK models, the scramblon calculations \cite{Gu:2021xaj,Zhang:2022fma} predict that the resolved metric $K_{mn}(\ell)$ takes the factorized form $K_{mn}(\ell) = \ell J_m(\ell) J_n(\ell)$.
The factorization indicates the following structure of the operator wave-function$|\mathcal{O}_n \rangle = \sum_{\ell} J_n(\ell)\;|\chi_\ell \rangle$.
i.e. the projection onto operator size $\ell$ is identical for all $\mathcal{O}_n$. We expect this as a result of the permutation symmetric Hamiltonians. Upon scaling the operator size as: $\ell^{1/\varkappa} \propto \lambda n$, we obtain the asymptotic behavior 
\begin{eqnarray}
    J_n(\ell) \sim n^{-\varkappa/2}\left[e^{\sqrt{\lambda(\lambda-4)}}\left(\lambda-2-\sqrt{\lambda(\lambda-4)}\right)\right]^{-n}
\end{eqnarray}
This implies that $\mathcal{O}_n$ has typical operator size $\ell \sim n^\varkappa$. The operator weight shows a phase transition between oscillatory behavior for $\lambda<4$ to exponential decay for $\lambda>4$. For the MBL systems, the resolved Krylov metric $K_{mn}(\ell)=K_{nn}(\ell) \delta_{mn}$ remains exactly diagonal. This indicates that the projections from distinct $\mathcal{O}_{m\neq n}$ onto any fixed operator length $\ell$ are orthogonal. Upon scaling the operator size as: $\ell \propto \lambda\ln{n} $, we obtain the asymptotic behavior 
\begin{equation}
    K_{nn}(\ell) \sim \lambda\; e^{-\frac{(\lambda-\xi)^2}{\xi}\ln{n}}
\end{equation}
This implies a Gaussian distribution in the operator size $\ell$ with comparable mean and variance: $\langle \ell \rangle \sim \delta^2 \sim  \xi \ln{n}  $.

 \emph{ \color{blue}Discussions.--} In this work, we introduced the Krylov metric $K_{mn}$ to bridge the gap between the growth of Krylov complexity and quantum many-body chaos. Physically, the Krylov metric measures the size growth in the Krylov space, which captures intrinsic properties of the Krylov basis. With a combination of the Krylov metric and Lanczos coefficients, we are able to provide criterion for fast scramblers, which require power-law growing diagonal components with negligible off-diagonal components for $K_{mn}$, in addition to exponential growing Krylov complexity. These criteria are supported by analytical studies in the SYK model, Luttinger liquids, and MBL systems.

 We conclude with some proposals for future investigations. Firstly, while our criterion in terms of the Krylov metric provide sufficient conditions for fast scramblers, it is important to understand to what extent are they also necessary. Secondly, having decomposed the lyapunov exponent $\varkappa = 2\alpha h$ into factors from the Krylov complexity $\alpha$ and the Krylov metric $h$, we need to understand the physical roles each factor plays. For example, how do they reflect different aspects of the underlying mechanism for quantum many-body chaos? Lastly, equipped with the perspective of the Krylov metric, we can explore systems exhibiting novel behaviors beyond the ones identified in this work. This will possibly shed lights on new aspects of quantum chaos.   


\textit{Acknowledgment.}
We thank Zhao-Yi Zeng and Ren Zhang for helpful discussions.
This project is supported by the NSFC under grant No. 12175238, 12374477, 12247103.

\bibliography{ref.bib}

\onecolumngrid
\begin{center}
\newpage\textbf{\large
Supplemental Material: Dissecting Quantum Many-body Chaos in the Krylov Space}
\\
\vspace{4mm}

\end{center}

 In this suppelmentary material, we provide additional details for the calculations done in the main text.

  \section{Asymptotic behaviors of $K_{mn}$} 
  The connected part of the OTOCs $F(t_1,t_2)$ at late times are controlled by the asymptotic behaviors of the corresponding Krylov metric $K_{mn}$, which are related by: 
  \begin{equation}\label{eq:SYK_basic}
  F(t_1,t_2) = \sum_{m,n} K_{mn}\;\varphi_m(t_1) \varphi_n(t_2)^*      
  \end{equation} 
The operator wave-functions $\varphi_n(t)$ depend on the auto-correlation function. In cases where they admit simple $n$-dependences, it is possible to perform the conversion explicitly. More explicitly, for $\varphi_n(t)$ of the general form: 
\begin{equation}
    \varphi_n(t) = D(n)h(t) y(t)^n 
\end{equation}
We can extract the Krylov by expanding $F(t_1,t_2)$ in powers of $y(t_1)$ and $y(t_2)$, which in turn can be written as double contour integrals: 
\begin{equation}
    K_{mn} = D(n)^{-1}D(m)^{-1} \oint \frac{dy_1}{y_1^{n+1}} \oint \frac{dy_2}{y_2^{m+1}} h(t_1)^{-1}h(t_2)^{-1}F(t_1,t_2)
\end{equation}
where the integrand can be viewed implicitly as functions of the $y_{1,2}=y(t_{1,2})$. For the interest of large order asymptotics $m,n\gg 1$, we can use $m,n$ as large parameters to perform saddle-point approximations for evaluating these contour integrals. In this section, we derive the asymptotic behaviors of $K_{mn}$ for the three classes of models considered in the main text. 

\subsection{Example 1: the SYK models}

We begin with the example of the SYK models. Quoting the expressions for $F(t_1,t_2)$ and $\varphi_n(t)$ in the main text gives:
\bea
F(t_1,t_2) &=& G\left(t_1+t_2\right)H\left(t_1-t_2\right),\;\;G(t)=e^{h\alpha t},\;\;H(t)=\cosh{(\alpha t)}^{-2\Delta-h}\nonumber\\
\varphi(t)&=&\frac{\tanh{(\alpha t)}^n}{\cosh{(\alpha t)}^{2\Delta}} D(n),\;\;D(n)=\sqrt{\frac{\Gamma(2\Delta+n)}{\Gamma(n+1)\Gamma(2\Delta)} }
\eea
In this case, we can obtain an exact analytic expression for $K_{mn}$ by identifying $y_{1,2} = \tanh{(\alpha t_{1,2})}$ and writing: 
\begin{equation}\label{eqn:tobeexpand}
\cosh{(\alpha t_1)}^{2\Delta}\cosh{(\alpha t_2)}^{2\Delta}F(t_1,t_2) \propto \frac{(1+y_1)^h(1+y_2)^h}{(1-y_1y_2)^{h+2\Delta}} = \sum_{mn} D_m D_n K_{mn} y_1^m y_2^n
\end{equation}
Here we fixed the normalization condition $K_{00}=1$. The Krylove metric $K_{mn}$ can be obtained by expanding each factor in the RHS of \eqref{eqn:tobeexpand} in terms of $y_{1,2}$:
\begin{equation}
(1+y)^h=\sum_{n=0}^\infty \frac{\Gamma(h+1) }{\Gamma(n+1)\Gamma(h+1-n)}y^n,\ \ \ \ \ \ \frac{1}{(1-y_1y_2)^{D}}=\sum_{n=0}^\infty \frac{\Gamma(n+D) }{\Gamma(n+1)\Gamma(D)} y_{1}^{n} y_{2}^{n}.
\end{equation}
Since $K_{mn}$ is symmetric in $m$ and $n$, we assume $m\geq n$. This gives: 
\begin{equation}
\begin{aligned}
K_{mn}=&\sqrt{\frac{\Gamma(n+1)\Gamma(2\Delta)}{\Gamma(2\Delta+n)} }\sqrt{\frac{\Gamma(m+1)\Gamma(2\Delta)}{\Gamma(2\Delta+m)} }\sum_{k=0}^n \frac{\Gamma(h+1) }{\Gamma(n-k+1)\Gamma(h+1-n+k)}\\&\times\frac{\Gamma(h+1) }{\Gamma(m-k+1)\Gamma(h+1-m+k)}\frac{\Gamma (h+k+2 \Delta )}{\Gamma (k+1) \Gamma (h+2 \Delta )}\\
=&\sqrt{\frac{\Gamma(2\Delta)}{\Gamma(n+1)\Gamma(2\Delta+n)} }\sqrt{\frac{\Gamma(2\Delta)}{\Gamma(m+1)\Gamma(2\Delta+m)} } \Gamma (h+1)^2\\&\times\frac{ \, _3F_2(-m,-n,h+2 \Delta ;h-m+1,h-n+1;1)}{\Gamma (h-m+1) \Gamma
   (h-n+1)}
\end{aligned}
\end{equation}
Here $\, _3F_2$ is the generalized hypergeometric function. In particular, for $m=n$, the result reduces to 
\begin{equation} \label{eq:SYK_exact}
\begin{aligned}
K_{nn}=\frac{\Gamma (2 \Delta ) \Gamma (h+n+2 \Delta ) \, _3F_2(-h,-h,-n;1,-h-n-2 \Delta
   +1;1)}{\Gamma (h+2 \Delta ) \Gamma (n+2 \Delta )}.
\end{aligned}
\end{equation}
Given the exact result (\ref{eq:SYK_exact}) for $K_{mn}$, we are more interested in its asymptotic behaviors. This can be more readily obtained via the contour integrals representation of $K_{mn}$ about $y_1=y_2=0$ and performing the suggested saddle-point analysis. We proceed by parameterizing both by their the phase variables $y_{1,2}=e^{\theta_{1,2}}$. The contour integral can thus be written as: 
\be
K_{mn} = D(m)^{-1}D(n)^{-1}\int d\theta_1 d\theta_2\; e^{-S(\theta_1,\theta_2)}
\ee
where the effective action $S$ is given by:
\bea
S\left(\theta_1,\theta_2\right) &=& m \theta_1 + n \theta_2 + (2\Delta+h)\ln{\left(1-e^{\theta_1+\theta_2}\right)}\nonumber\\
&-&h\ln{\left(1+e^{\theta_1}\right)}-h\ln{(1+e^{\theta_2})}
\eea
Treating $m,n\gg 1$ as the large parameters, the saddle-point equations becomes: 
\bea
m=\frac{h e^{\theta_1}}{1+e^{\theta_1}}+\frac{(2\Delta+h)e^{\theta_1+\theta_2}}{1-e^{\theta_1+\theta_2}},\;\;\;n=\frac{h e^{\theta_1}}{1+e^{\theta_2}}+\frac{(2\Delta+h)e^{\theta_1+\theta_2}}{1-e^{\theta_1+\theta_2}}
\eea
The solution takes the form: 
\bea\label{eq:exact_saddle}
e^{\theta^*_1} &=& -\frac{h^2+2(n-m)(m+\Delta)\pm \sqrt{h^4+4(n-m)^2\Delta^2+4h^2\left(mn+(m+n)\Delta\right)}}{2(h+n-m)(m+2\Delta)} \nonumber\\
e^{\theta^*_2} &=& -\frac{h^2+2(m-n)(n+\Delta)\pm \sqrt{h^4+4(n-m)^2\Delta^2+4h^2\left(mn+(m+n)\Delta\right)}}{2(h+m-n)(n+2\Delta)}\nonumber
\eea
We analyze the properties of the saddle in different limits of $M,N\gg 1$. 
\begin{itemize}
\item \textbf{$\boldsymbol{n=L(1+\lambda),\;m=L(1-\lambda),\;\;0<\lambda<1}$:}

In this limit, the entry is away from the diagonal by the same order as $m,n$, and the parameter $\lambda$ denotes the orthogonal distance to the diagonal. The dominant saddle point admits an expansion in large $L$: 
\bea
\theta_1^*&=& i\pi -\frac{2\Delta \lambda+h(1-\lambda)-\sqrt{h^2(1-\lambda^2)+4\Delta \lambda^2}}{2L\lambda(1-\lambda)}+...\nonumber\\
\theta_2^*&=& i\pi -\frac{2\Delta \lambda-h(1+\lambda)+\sqrt{h^2(1-\lambda^2)+4\Delta \lambda^2}}{2L\lambda(1+\lambda)}+...
\eea
Plugging this back to the effective action gives, we can estimate the large order behavior of the Krylov kernel: 
\be
K_{mn} \sim  L^{-h+1}\lambda^{-2h} \times \left(\text{fluctuation}\right)
\ee
where we have kept only the leading order dependence on finite but small $\lambda$, The fluctuation part comes from the integral about the saddle point. Expanding near the $(\theta_1^*, \theta^*_2)$, it can be checked that the effective action takes the form: 
\bea
S(\theta_1,\theta_2) = S^* + A_+ L^2 \delta \theta_+^2+A_- L^2 \lambda^2 \delta\theta_-^2+\sum_{p+q\geq 3} S_{pq}\;\delta \theta^p_+\; \delta \theta^q_-,\;\; S_{pq}\sim L^{p+q}\lambda^q\nonumber
\eea 
where $A_{\pm}$ are $\mathcal{O}(1)$ constants and $\delta \theta_{\pm}$ are eigen-modes of the Hessian matrix. At the leading order in $L\gg 1$ they simply correspond to: 
\be 
\delta \theta_+ = \delta \theta_1+\delta \theta_2,\;\;\delta \theta_+ = \delta \theta_1-\delta \theta_2,\;\;\delta \theta_{1,2} = \theta_{1,2}-\theta^*_{1,2}
\ee
We can therefore extract an additional factor of $L^{-2}\lambda^{-1}$ from the fluctuation by rescaling the integration variables $\delta \theta_+=\delta \tilde{\theta}_+/L, \delta \theta_- = \delta\tilde{\theta}_-/(L\lambda)$, and leaving the remaining integral as an order 1 factor: 
\be
\int^{\pi}_{-\pi} d\delta\theta_+\int^\pi_{-\pi} d\delta\theta_-\; e^{-\sum_{p,q} S_{pq}\; \delta \theta_+^p \delta \theta^q_-} = \frac{1}{L^2\lambda}   \left(\int^\infty_{-\infty}d \delta\tilde{\theta}_+ \int^\infty_{-\infty}d \delta\tilde{\theta}_-\; e^{-\sum_{p,q} \tilde{S}_{pq}\; \delta \tilde{\theta}_+^p \delta \tilde{\theta}^q_-}\right)\nonumber
\ee 
where we now have $\tilde{S}_{pq} \sim \mathcal{O}(1)$  and the integration range of $\delta \tilde{\theta}_{\pm}$ has been set to $\pm\infty$ after the rescaling. We remark that this is slightly different from the usual scenario in performing the saddle-point analysis, in the sense that the fluctuations are not weakly coupled as Gaussians, yet whose contribution one can extract as a scaling factor comparable to $S^*$. Combining these we arrive at the estimates: 
\be
K_{mn} \sim L^{-h-1}\lambda^{-2h-1}
\ee

\item \textbf{$\boldsymbol{n=m=L}$:}

They correspond to matrix elements are lie exactly along the diagonal. The dominant saddle point now admits the expansion in large $L$: 
\be
\theta^*_1 = \theta^*_2 = -\frac{h+2\Delta}{2L}+...  
\ee
Similar to before, this saddle-point evaluates to the following estimate: 
\be
K_{nn} \sim n^{h+1}\times \left(\text{fluctuation}\right) 
\ee 
In this case, the fluctation analysis proceeds slightly differently. The expansion of the effective action now takes the form: 
\be
S(y_1,y_2) = S^* + \sum_{p,q} S_{pq} \delta \theta_+^p \delta \theta _-^q,\;\;\;S_{pq} \sim L^p
\ee
We see that in contrary to the previous case, the expansion coefficients scale with only one of the modes, i.e. $\delta \theta_+$. By the same logic as before, we can rescale $d\theta_+=d\tilde{\theta}_+/L$, and after doing this we should extract an additional factor of $L^{-1}$ from the fluctuation factor. As a result the diagonal Krylov kernel now exhibits the expected scaling behavior: 
\be
K_{nn} \sim n^{h}
\ee
\end{itemize}

\begin{figure}[t]
\centering
\includegraphics[width=0.90\linewidth]{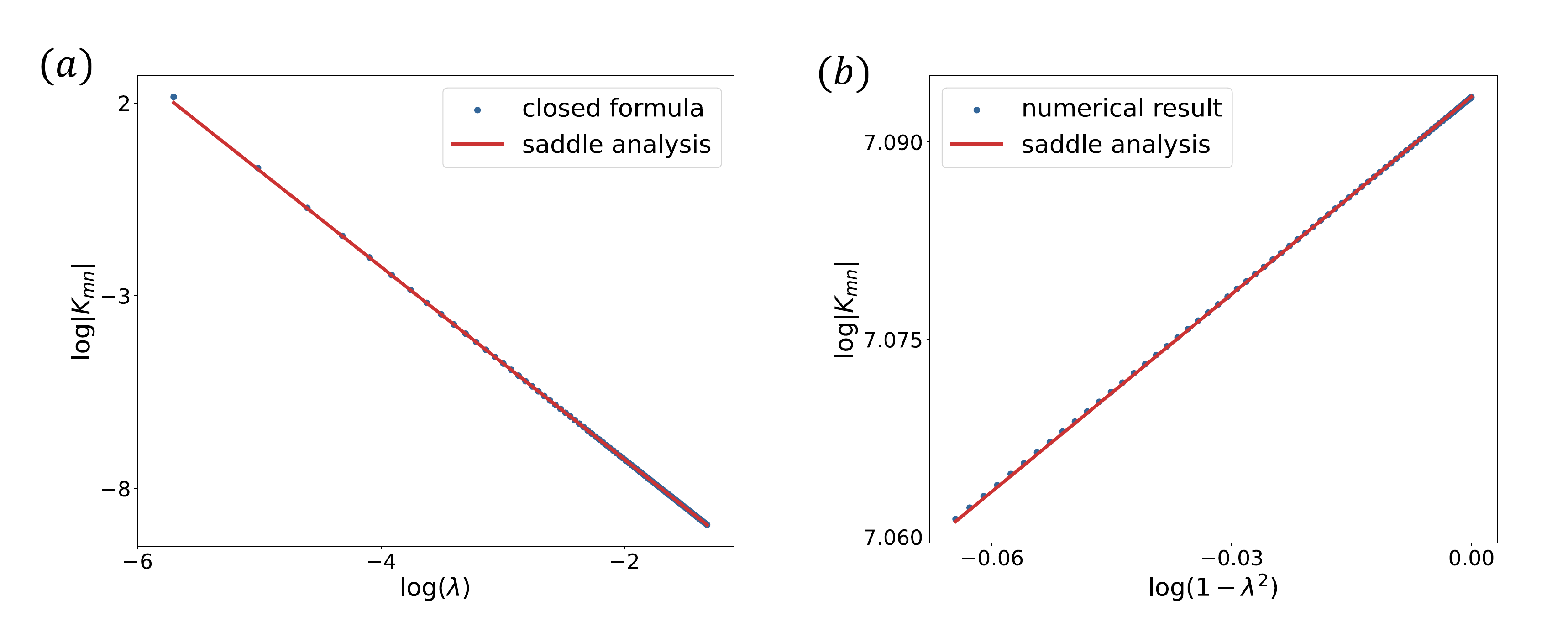}
\caption{We present the asymptotic behaviors of the orthogonal off-diagonal  Krylov metric $K_{mn}$ with $m=L(1-\lambda), n=L(1+\lambda)$ for: (a) the SYK model with $h=3/4$, (b) the Luttinger liquids with $\Delta=1$. In both cases, we have chosen $L = 300 $. }
\label{fig:nonDiagonalKmn}
\end{figure}

Combining both limits, we summarize the asymptotic behaviors of $K_{mn}$ as follows. The diagonal elements are given by: 
\be
K_{nn}\sim n^{h}
\ee
while the off-diagonal elements satisfy:
\be 
K_{mn} \sim K_{LL}|m-n|^{-2h-1},\;\;L = \frac{m+n}{2}
\ee
We also verify this off-diagonal result by doing numerical calculation directly, see FIG. \ref{fig:nonDiagonalKmn} (a). In other words, the off-diagonal elements decay along the orthogonal off-diagonal direction as a power law with power $(-2h-1)$. This is sufficient for the dominance of the diagonal contribution for general $h$. In addition, it is also observed that the relative decay becomes singular for the immediate off-diagonal matrix elements as $h\to \lbrace 0,1\rbrace$: 
\be
\frac{K_{n,n+1}}{K_{nn}}\sim h(1-h),\;\; \frac{K_{n,m}} {K_{n,m-1}} \sim \mathcal{O}(1),\; \text{for } (m-n)\geq 2
\ee
Therefore, in the either the dissipative limit $h\to 0$ or the maximally chaotic limit $h\to 1$, the Krylov metric approaches being exactly diagonal. Capturing such phenomena however is beyond the scope of saddle-point approximations. 

\subsection{Example 2: Luttinger liquids}
We begin the example of luttinger liquids by writing down the Hamiltonian as: 
 \begin{equation}
     \begin{aligned}
         H &= \frac{u}{2 \pi} \int d x \left[ \frac{1}{K} \left(\nabla \phi(x)\right)^2 + K \left(\pi \Pi(x)\right)^2\right] \\
         &=  \frac{u}{2 \pi} \int d x \left[ \frac{1}{K}\left(\nabla \phi(x)\right)^2 + K \left(\nabla \theta(x)\right)^2\right] \\
     \end{aligned}
 \end{equation}
 where $ \nabla \theta(x)/\pi =  \Pi(x)$ is the canonically conjugate momentum  of $\phi(x) $,  and the commutation relation can be written as:
 \begin{equation}
     \left[\phi\left(x\right), \frac{1}{\pi} \nabla \theta(x^{\prime})\right]= i \delta\left(x-x^{\prime}\right)
 \end{equation}
One can then derive the following finite temperature correlation functions between general vertex operators
\cite{10.1093/acprof:oso/9780198525004.001.0001}
\bea\label{eq:luttinger_general}
I &=& \left\langle\prod_j e^{i A_j \phi\left(r_j\right)}\right\rangle_{\beta}= e^{\frac{1}{2} \sum_{i<j}\left[\left(A_i A_j K \right) F\left(r_{i}-r_{j}\right)\right]}\nonumber\\
 F(r)  &=& \frac{1}{2} \log \left[\sinh ^2\left(\frac{\pi x}{\beta u}\right)+\sin ^2\left(\frac{\pi \tau}{\beta}\right)\right]
\eea
Consider the vertex operator of scaling dimension $\Delta  = K n^{2}/4$:
\begin{equation}
    V_{n}(x,t) = :\exp( i n \phi(x,t)):
\end{equation}
Let us compute the thermally regulated OTOC of this operator:
\begin{equation}
    \begin{aligned}
        C(t_1, t_2) = \left\langle V_{-n}(t_{1} -i 3\beta/4) V_{n}(-i\beta/2) V_{n}(t_2 -i \beta/4) V_{-n}(0)\right\rangle_{\beta}
    \end{aligned}
\end{equation}
Applying the general formula \eqref{eq:luttinger_general} to this computation then corresponds to setting the following non-zero parameters:
\begin{equation}
    A_{1} = -n ,\quad A_{2} = n,\quad A_{3}=n, \quad A_{4} = -n
\end{equation}
and the corresponding coordinates are only separated in the time direction:
\begin{equation}
    \begin{aligned}
        r_{1} = \left(0, -i u(t_{1} -i 3\beta/4)\right), \quad  r_{2} = \left(0, -i u(-i \beta/2)\right),\quad r_{3} = \left(0, -i u (t_{2} -i\beta/4)\right), \quad  r_{4} = (0,0)
    \end{aligned}
\end{equation}
The normalized OTOC is then given by: 
\begin{equation}
    \begin{aligned}
        C(t_1, t_2)  &= e^{\frac{K}{2} \sum_{i<j}\left(A_i A_j  F_1\left(r_{i}-r_{j}\right)\right)} = \exp \left( \frac{K n^{2}}{2} W(t_1,t_2)\right), \\
        W(t_1, t_2) &= -F(r_1 - r_2) - F(r_1 - r_3) +  F(r_1 - r_4) + F(r_2-r_3) -F(r_2- r_4) - F(r_3-r_4)\nonumber\\
        & = \frac{1}{2} \log \left[\frac{\sinh ^2\left(\frac{\pi }{\beta} (t_1 - i 3\beta/4)\right) \sinh ^2\left(\frac{\pi }{\beta} (t_2 + i \beta/4)\right)}{\sinh ^2\left(\frac{\pi }{\beta} (t_1 - i \beta/4)\right) \sinh ^2\left(\frac{\pi }{\beta} (t_1- t_2 - i \beta/2)\right) (-i)^{2} \sinh ^2\left(\frac{\pi }{\beta} (t_2 - i \beta/4)\right)}\right]\nonumber
    \end{aligned}
\end{equation}
In the end, we obtain the following explicit form:
\begin{equation}
    C(t_1,t_2) = \left[ \frac{\cosh\left(\frac{\pi }{\beta}(t_1 -t_2)\right) -i \sinh\left(\frac{\pi }{\beta} (t_1+ t_2)\right)}{\left(\cosh\left(\frac{\pi }{\beta}(t_1 -t_2)\right) +i \sinh\left(\frac{\pi }{\beta} (t_1+ t_2)\right)\right) \cosh \left(\frac{\pi }{\beta} (t_1- t_2)\right) } \right]^{ 2 \Delta }\nonumber
\end{equation}
Based on these, we can now work out the corresponding Krylov metric. Being a CFT, the finite temperature auto-correlation function of the Luttinger liquids takes the same form as that of the SYK models. As a result, the operator wavefunction in the Krylov basis is identical to (\ref{eq:SYK_basic}), and we have that: 
\be
\sum_{m,n}\varphi_m(t_1) \varphi_n(t_2) K_{mn} = F(t_1,t_2) = \cosh{\left(\alpha t_{12}\right)}^{-2\Delta}\left(1- \left[ \frac{\cosh\left(\alpha t_{12}\right) -i \sinh\left(2\alpha T_{12}\right)}{\cosh\left(\alpha t_{12}\right) +i \sinh\left(2\alpha T_{12}\right) } \right]^{ 2 \Delta }\right).
\ee
where we have defined $\alpha = \pi /\beta, t_{12}=t_1-t_2, T_{12}=t_1+t_2$. The Krylov metric can therefore be computed by performing a similar double contour integral:
\bea 
K_{mn} &=& \delta_{mn}- D(m)^{-1}D(n)^{-1} \oint \frac{dy_1}{y_1^{m+1}} \oint \frac{dy_2}{y_2^{n+1}}  \left[\frac{(1-iy_1)(1-iy_2)}{(1+iy_1)(1+iy_2)(1-y_1y_2)}\right]^{2\Delta}\nonumber\\
&=&\delta_{mn}- D(m)^{-1}D(n)^{-1}  \int d\theta_1 d\theta_2\;e^{-S(\theta_1,\theta_2)}\nonumber
\eea
The effective action and saddle-point equation is given by:
\bea
&& S(\theta_1,\theta_2) =   m \theta_1 +  n \theta_2 -2 \Delta \log \left[\frac{\left(1-i e^{\theta_1}\right)\left(1-i e^{\theta_2}\right)}{\left(1+i e^{\theta_1}\right)\left(1+i e^{\theta_2}\right)\left(1-e^{\theta_1+\theta_2}\right)}\right] \nonumber\\
&& m +\frac{2i\Delta }{\cosh{\theta_1}}+\frac{2\Delta}{1-e^{-(\theta_1+\theta_2)}}=0,\;\;n +\frac{2i\Delta }{\cosh{\theta_2}}+\frac{2\Delta}{1-e^{-(\theta_1+\theta_2)}}=0
\eea
These are high degree polynomial equations of $y_{1,2} =e^{\theta_{1,2}}$. In the limit of: 
\be
m=L(1-\lambda),\;\;n=L(1+\lambda),\;\; L\gg 1
\ee
The dominant saddle-point can be obtained in series expansion of $L^{-1}$: 
\bea 
\theta^*_1 = -\frac{\pi i}{2}+\frac{2i\Delta}{1-\lambda}L^{-1}+\mathcal{O}(L^{-2})\nonumber\\
\theta^*_2 = -\frac{\pi i}{2}+\frac{2i\Delta}{1+\lambda}L^{-1}+\mathcal{O}(L^{-2})
\eea
Plugging this into the effective action, and neglecting the subdominant $\delta_{mn}$ term in $K_{mn}$ then gives: 
\be
K_{mn} \sim (-1)^{L} L^{2\Delta+1}(1-\lambda^2)^{\Delta+1/2}\times \text{(fluctuation)}
\ee
Again, this result is matched well with the numerical result, for the diagonal part one can see the figures in maintext, for the off-diagonal, see FIG \ref{fig:nonDiagonalKmn} (b). The alternating sign factor $(-1)^L$ comes from the imaginary leading order terms of $\theta^*_{1,2}$. The fluctuation part of the integral produces an additional factor $L^{-2}(1-\lambda^2)^{-1}$, which can be revealed by a similar analysis as before. We omit the details. Combining these factors we obtain that: 
\be
K_{mn} \sim (-1)^{L} L^{2\Delta-1}(1-\lambda^2)^{\Delta-1/2} = (-1)^{\frac{m+n}{2}} (mn)^{\Delta -1/2}
\ee

\subsection{Example 3: MBL systems}
The effective Hamiltonian for MBL systems takes the form: 
\be
 H = \frac{1}{2} \sum_{i\neq j} J_{ij} \sigma^i_z \sigma^j_z + \sum_i h_i \sigma^i_z,  \quad i, j \in [-N/2+1,N/2],  \quad \text{with }N \to \infty
\ee
where the coefficients $\lbrace J_{ij}, h_i\rbrace$ are independent Gaussian random variables with: 
\be
\langle J_{ij}\rangle = 0,\;\;\langle J_{ij}^2\rangle = J^2 \exp\left(\frac{|i-j|}{\xi}\right),\;\;\langle h_i\rangle = 0,\;\;\langle h_i^2\rangle = h^2
\ee
We study the time evolution of the following operator localized at site $i=0$: 
\be
 \sigma^0_x(t) = e^{iHt}\;\sigma^0_x\; e^{-iHt}
\ee
and construct the Krylov basis using the infinite temperature operator norm: 
\be
\langle A, B\rangle = \text{Tr} \left( A^\dagger B\right) 
\ee 
The lanzcos coeffcients are then fixed by the infintie temperature auto-correlation function, which we now compute.
\bea
C(t)=\langle \sigma^0_x(t),\sigma^0_x(0)\rangle = \text{Tr} \left(e^{iHt} \sigma^0_x e^{-iHt} \sigma^0_x\right) = \text{Tr}\left(\exp\left[2it\sum_{j\neq 0}{J_{0j}}\sigma^0_z \sigma^j_z + 2it h_0  \sigma^0_z\right]\right)\nonumber
\eea
where we have used that: 
\be\label{eq:pauli_idenity_1}
\sigma^0_x \sigma^j_z \sigma^0_x = -\delta_{j0}\; \sigma^j_z\;\rightarrow \sigma^0_x\;e^{-iHt}\;\sigma^0_x = e^{-iHt} \times \exp\left[2it\sum_{j\neq 0}{J_{0j}}\sigma^0_z \sigma^j_z + 2it h_0 \sigma^0_z\right]
\ee
Applying the identity $e^{iJ \sigma_z} = \cos{(J)}+i\sin{(J)}\sigma_z$, the trace can be easily evaluated: 
\bea
C(t)&=&\text{Tr} \prod_{j\neq 0} \left(\cos{(2J_{0j}t)}+i\sin{(2J_{0j}t)\sigma^0_z \sigma^{j}_z}\right)\left(\cos{(2h_0 t)}+i\sin{(2h_0t)}\sigma^0_z\right)\nonumber\\
&=& \prod_{j\neq 0} \cos{(2J_{0j}t)} \cos{(2h_0 t)}
\eea
A more explicitly expression can be obtained by taking the statistical average: 
\bea 
\overline{C(t)} =\prod_{j\neq 0} \overline{\cos{(2J_{0j}t)}}\; \overline{\cos{(2h_0 t)}}= e^{-\frac{\gamma^2}{2}t^2},\;\;\gamma^2 =4J^2 \sum_{j\neq 0} e^{-|j|/\xi} + 4 h^2
\eea
We see that the auto-correlation function decays like a Gaussian, which happens to also be the case where the krylov basis wave-function and the Lanzcos coefficients are known explicitly: \cite{Caputa:2021sib}: 
\be\label{eq:gaussian_krylov}
C(t)=e^{-\frac{\gamma^2 t^2}{2}} \;\;\rightarrow\;\; \varphi_n(t) = e^{-\frac{\gamma^2 t^2}{2}} \frac{\gamma^n t^n}{\sqrt{n!}},\;\;b_n = \gamma \sqrt{n}
\ee
In particular, for MBL systems the Lanzcos coefficients grow sub-linearly as $\sqrt{n}$. Notice that $\varphi_n(t)$ is also of the form that allows extracting the Krylov metric from the OTOC via explicit contour integrals. We start by considering:
\bea
F(t_1,t_2)&=&-\frac{1}{N}\sum_{m}\text{Tr}\left[\sigma^0_x(t_1),\sigma^m_x\right]\left[\sigma^0_x(t_2),\sigma^m_x\right]\nonumber\\
&=&2 C(t_{12})-\frac{2}{N}\sum_m\text{Tr}\left[\sigma^0_x(t_1)\; \sigma^m_x\; \sigma^0_x(t_2)\;\sigma^m_x\right]
\eea
where $N$ is total number of sites in the Hamiltonian. The OTOC term can be evaluated via similar tricks: 
\bea
&&OTOC =  \sum_{m}\text{Tr}\left[e^{i H t_1}\sigma^0_x e^{-iH t_1}\; \sigma^m_x\; e^{iHt_2}\sigma^0_x e^{-iHt_2}\;\sigma^m_x\right]\nonumber\\
&=&  \sum_m\text{Tr}\Big[e^{2it_1\sum_{j\neq 0}{J_{0j}}\sigma^0_z \sigma^j_z + 2it_1 h_0 \sigma^0_z}\;\sigma^0_x \sigma^m_x\; e^{2it_2\sum_{j\neq 0}{J_{0j}}\sigma^0_z \sigma^j_z + 2it_2 h_0 \sigma^0_z}\;\sigma^0_x \sigma^m_x\Big]\nonumber\\
&=& C(t_1+t_2) + \sum_{m\neq 0} \text{Tr}\Big[e^{2it_1\sum_{j\neq 0}{J_{0j}}\sigma^0_z \sigma^j_z + 2it_1 h_0 \sigma^0_z}\sigma^m_x\; e^{-2it_2\sum_{j\neq 0}{J_{0j}}\sigma^0_z \sigma^j_z - 2it_2 h_0 \sigma^0_z}\;\sigma^m_x\Big]\nonumber\\
&=&C(t_1+t_2) + \sum_{m\neq 0} \cos{\left(2J_{0m} (t_1+t_2)\right)} \prod_{j\neq 0,m} \cos{\left(2J_{0j}t_{12}\right)}\cos{(2h_0(t_{12}))}
\eea
Taking the statistical average then gives: 
\be\label{eq:OTOC_MBL}
\overline{F(t_1,t_2)} = 2 e^{-\frac{\gamma^2}{2} t^2_{12}} -\frac{2}{N} e^{-\frac{\gamma^2}{2} (t_1+t_2)^2}-\frac{2}{N}\sum_{m\neq 0} \exp\left(-8 J^2 e^{-|m|/\xi} t_1 t_2\right)e^{-\frac{\gamma^2}{2}\; t_{12}^2} 
\ee
The explicit dependence on $\lbrace t_1,t_2 \rbrace$ in (\ref{eq:OTOC_MBL}) is complicated through the summation. In the late time limit $J^2 t_1 t_2 /\xi \gg 1$, we can approximate the summation by replacing all those terms with small exponents, i.e. $8J^2 e^{-|m|/\xi} t_1t_2 \leq 1$, by $1$; and the others by $0$. Doing this then gives:  
\be
\sum_{m\neq 0} \exp{\left(-8 J^2 e^{-|m|/\xi} t_1 t_2\right)} \approx (N-1)- 2\xi\log{\left(8J^2t_1 t_2\right)}
\ee
Plugging this back, we therefore obtain that: 
\be\label{eq:MBL_OTOC}
\overline{F(t_1,t_2)} \approx  \frac{2}{N}e^{-\frac{\gamma^2}{2}(t_1-t_2)^2}-\frac{2}{N}e^{-\frac{\gamma^2}{2} (t_1+t_2)^2}+\frac{2}{N} \xi\ln{\left(8J^2 t_1t_2\right)}e^{-\frac{\gamma^2}{2}\; (t_1-t_2)^2} 
\ee
Using the form of the operator wavefunction (\ref{eq:gaussian_krylov}), the asymptotic Krylov metric is then related to $\overline{F(t_1,t_2)}$ at large $t_1, t_2$ via: 
\be 
\sum_{m,n}t_1^m t_2^n K_{mn} = \frac{\sqrt{m!n!}}{\gamma^{m+n}} e^{\frac{\gamma^2}{2}(t_1^2+t_2^2)}\;\overline{F(t_1,t_2)} \approx \frac{2\sqrt{m!n!}}{N\gamma^{m+n}} \left(e^{\gamma^2 t_1 t_2}-e^{-\gamma^2 t_1 t_2} +2\xi \ln{\left(8J^2 t_1 t_2\right)}e^{\gamma^2 t_1 t_2}\right)
\ee
The RHS only depends on the product $(t_1t_2)$. As a consequence, the asymptotic Krylov metric is diaogonal: 
\be
K_{mn} =K_{nn} \delta_{mn}
\ee
The diagonal elements can therefore be extracted by a single contour integral in $x=\gamma^2 t_1 t_2$:
\bea
K_{n} \approx  \frac{2}{N}-\frac{2(-1)^n}{N} + \frac{2\xi n!}{N}\oint \frac{dx}{x^{n+1}} \ln{\left(\frac{8J^2}{\gamma^2}x\right)} e^{x}
\eea
At large order $n\gg 1$ the integral is approximated by the contribution from the dominant saddle-point $x^*$ satisfying: 
\be 
n+1 = x^*+\ln{\left(\frac{8J^2}{\gamma^2}x^*\right)}^{-1}\rightarrow x^* \approx n
\ee
The fluctuation about the saddle gives an additional $\sqrt{n}$ factor. Combining these then gives the asymptotic behavior:
\be 
K_{nn} \approx \frac{2}{N}-\frac{2(-1)^n}{N}+\frac{2\xi}{N}\ln{\left(\frac{8J^2}{\gamma^2} n\right)} \approx \frac{2\xi}{N}\ln{\left(\frac{8J^2}{\gamma^2} n\right)}
\ee

\section{Size-resolved Krylov metric}
The OTOCs can often be interpreted in terms of the operator-spreading under time-evolution. In models where the operator-size can be explicitly defined, one can construct eigen-states in the operator space with fixed operator-size: 
\be
\hat{N}|\mathcal{O}_n \rangle = n |\mathcal{O}_n\rangle
\ee
where $\hat{N}$ is the super-operator that measures the size of the operator states. For example, in the SYK models and the MBL systems the eigen-states of size $n$ consists of:
\be
|\chi_{i_1}...\chi_{i_n}\rangle,\;\;\; |\sigma^{i_1}_{\alpha_1}...\sigma^{i_n}_{\alpha_n}\rangle,\;\;\alpha_i \in \lbrace x,y,z\rbrace
\ee
In these models, $F(t_1,t_2)$ and the Krylov metric $K_{mn}$ can be interpretted as the matrix elements of $\hat{N}$: 
\be 
F(t_1,t_2) = \langle \mathcal{O}(t_1)|\;\hat{N}\;|\mathcal{O}(t_2)\rangle,\;\;K_{mn} = \langle \mathcal{O}_m |\;\hat{N}\;|\mathcal{O}_n\rangle
\ee
We can probe more refined aspects of the Krylov metric by further resolving it into contributions from the operator space sectors, each of which contains operator states of only fixed sizes: 
\be
K_{mn} = \sum_{\ell} K_{mn}(\ell),\;\;\; K_{mn}(\ell) = \langle\mathcal{O}_m |\; \hat{N} \hat{P}(\ell)\;|\mathcal{O}_n\rangle = \ell\; \hat{P}(\ell)_{mn}
\ee
where $\hat{P}$ is the super-projector into operator space with fixed operator size $\ell$. The operator-size distribution $\hat{P}(\ell)_{mn}$ is equivalently encoded in the generating function, which is more accessible by explicit computation: 
\be
Z(t_1,t_2,\mu) = \langle \mathcal{O}(t_1)|\;e^{-\mu \hat{N}}\;|\mathcal{O}(t_2) \rangle =  \sum_{\ell,m,n} e^{-\ell \mu} \hat{P}(\ell)_{mn}\; \varphi_m(t_1)\; \varphi_n(t_2)^*
\ee
Therefore, by computing the operator-size generating function $Z(t_1,t_2,\mu)$, we can retrieve the distribution $\hat{P}(\ell)_{mn}$ and hence the resolved Krylov metric $K_{mn}(\ell)$. In this section, we will perform this calculation for the examples where the operator-size can be explicitly defined,  i.e. the SYK models and the MBL systems.   

\subsection{Example 1: the SYK models}
We begin with the SYK models considered in the main text. Recall that these models contain $N$ majorana fermions $\chi_i,i=1,...,N$ satisfying the anti-commutation relations: 
\be
\lbrace \chi_i, \chi_j\rbrace = 2\delta_{ij} 
\ee 
The generating function $Z(t_1,t_2,\mu)$ can be computed more conveniently by working in the doubled Hilbert-space $H_L\otimes H_R$. This allows us to defining the size super-operator explicitly as: 
\be
\hat{N} =\sum_i \frac{1}{2}\left(1+i\chi^L_i \chi^R_i \right)
\ee   
where $\chi^L_i$ and $\chi^R_i$ are majorana fermions acting on $H_L$ and $H_R$ respectively satsifying $\lbrace \chi^L_i,\chi^R_j\rbrace = 0$. To be compatible with this definition, the operator norm $\langle .\rangle$ is defined by the expectation value: 
\be
\langle \alpha,\beta\rangle = \langle I|\;\alpha^{\dagger}_L \beta_L\; |I\rangle = \text{Tr} \left(\alpha^\dagger \beta\right)
\ee
in the maximally entangled state $|I\rangle$: 
\be
|I\rangle \propto \Pi^N_{i=1} c^\dagger_i\; |\Omega\rangle,\;\;\;c_i=\frac{1}{2}\left(\chi^L_i-i \chi^R_i\right)
\ee 
where $|\Omega\rangle=\Pi^N_{i=1}|\Omega\rangle^L_i \otimes |\Omega\rangle^R_i$ is the product state of the fermionic vacua for all $\chi^{L,R}_i$. The state $|I\rangle$ is chosen so that it is annihilated by the combination of fermion operators: 
\be\label{eq:I_identity}
\left(\chi^L_i+i\chi^R_i\right) |I\rangle = 2c^\dagger_i |I\rangle= 0 
\ee
It is then easy to check that: 
\be
\hat{N}\; \chi^L_{i_1} ... \chi^{L}_{i_n} |I\rangle = n\; \chi^L_{i_1} ... \chi^{L}_{in} |I\rangle 
\ee
and therefore fulfilling definition of the operator size operator. The definitions of the operator size can be extended to finite temperatures by replacing $|I\rangle$ by a thermal field double state $|TFD\rangle$: 
\be
|TFD\rangle \propto e^{-\frac{\beta}{4}\left(\hat{H}_L+\hat{H}_R\right)}|I\rangle
\ee
Analogous to before, in computing the norm using $|TFD\rangle$ we also separate the two operators by a $\rho^{1/2}$ insertion: 
\bea
\langle \psi,\gamma_2\rangle_{TFD} &=&  \langle I|e^{-\frac{\beta}{8}\left(\hat{H}_L+\hat{H}_R\right)}\psi^\dagger_L e^{-\frac{\beta}{4}\hat{H}_L} e^{-\frac{\beta}{4}\hat{H}_L}\gamma_L e^{-\frac{\beta}{8}\left(\hat{H}_L+\hat{H}_R\right)}|I\rangle \nonumber\\
&=& \text{Tr} \left(\rho^{1/4} \psi^{\dagger} \rho^{1/2} \gamma\rho^{1/4}\right),\;\;\;\rho= e^{-\beta \hat{H}}
\eea
Alternatively this can can be understood as measuring the original operator size in the operator ``smeared" by the thermal density matrix: $O\to \rho^{1/4} O \rho^{1/4}$.

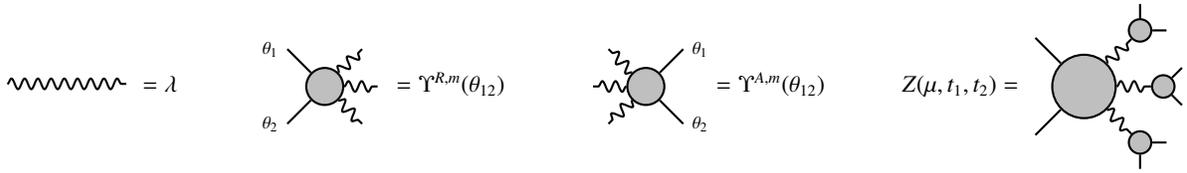
\begin{figure}[t]\centering
    \begin{tikzpicture}[scale=1.5,baseline={([yshift=-3pt]current bounding box.center)}]

\draw[thick,wavy] (0pt,0pt) to (30pt,0pt);

\filldraw  (0pt,0pt) circle (0pt) node[left]{ };
\filldraw  (30pt,0pt) circle (0pt) node[right]{ };
\end{tikzpicture}$=\lambda$
\ \  \ \ \ \ \ \  \ \ \ \ 
\begin{tikzpicture}[scale=1.25]
\node[bvertexnormal] (R) at (-30pt,0pt) {};
\draw[thick] (R) -- ++(135:16pt) node[left]{\scriptsize $\theta_1$};
\draw[thick] (R) -- ++(-135:16pt) node[left]{\scriptsize $\theta_2$};
\draw[wavy] (R) -- ++(45:16pt) node[left]{};
\draw[wavy] (R) -- ++(-45:16pt) node[left]{};
\draw[wavy] (R) -- ++(0:16pt) node[left]{};
\end{tikzpicture}$\ \ \ =\Upsilon^{R,m}(\theta_{12})$\ \  \ \ \ \ \ \  \ \ \ \ 
\begin{tikzpicture}[scale=1.25]
\node[bvertexnormal] (A) at (30pt,0pt) {};
\draw[thick] (A) -- ++(45:16pt) node[right]{\scriptsize $\theta_1$};
\draw[thick] (A) -- ++(-45:16pt) node[right]{\scriptsize $\theta_2$};
\draw[wavy] (A) -- ++(135:16pt) node[left]{};
\draw[wavy] (A) -- ++(-135:16pt) node[left]{};
\draw[wavy] (A) -- ++(180:16pt) node[left]{};
\end{tikzpicture}$=\Upsilon^{A,m}(\theta_{12})$
\ \  \ \ \ \ \ \  \ \ \ \ 
$Z(\mu,t_1,t_2)=$\begin{tikzpicture}
\node[bvertex] (R) at (-0pt,0pt) {};
\node[bvertexsmall] (A1) at (30/1.4142pt,30/1.4142pt) {};
\node[bvertexsmall] (A2) at (30pt,0pt) {};
\node[bvertexsmall] (A3) at (30/1.4142pt,-30/1.4142pt) {};

\draw[thick] (R) -- ++(135:26pt) node[left]{\scriptsize$ $};
\draw[thick] (R) -- ++(-135:26pt) node[left]{\scriptsize$ $};
\draw[thick] (A1) -- ++(45+45:10pt) node{};
\draw[thick] (A1) -- ++(-45+45:10pt) node{};
\draw[thick] (A2) -- ++(45:10pt) node{};
\draw[thick] (A2) -- ++(-45:10pt) node{};
\draw[thick] (A3) -- ++(45-45:10pt) node{};
\draw[thick] (A3) -- ++(-45-45:10pt) node{};

\draw[wavy] (R) to (A1);
\draw[wavy] (R) to (A2);
\draw[wavy] (R) to (A3);
\end{tikzpicture}
\caption{ A sketch of diagrammatics in the scramblon effective theory.
}
\label{fig:scramblon}
\end{figure}

It is very difficult to compute the generating function $Z(t_1,t_2,\mu)$ from first principles. Fortunately for the SKY models, progress can be made using the effective description based on the scramblon mode, see Figure (\ref{fig:scramblon}). In this framework, the relevant dynamics can be captured by two ingredents: (i) the scramblon mode propagator $\lambda$, characterized by the lyapunov exponent $\varkappa$; (ii) the retarded/advanced vertex functions $\Upsilon^{R/A,m}(\theta_{12})$ describing the couplings to $m$ scamblon modes, where $\theta_{ij}=\theta_i-\theta_j$ for the complexified time of insertion $\theta=\frac{2\pi}{\beta}\left(\tau+it\right)$. For our purpose, we can set $\Upsilon^R_m=\Upsilon^A_m = \Upsilon_m$ by assuming time-reflection symmetry. For more details please refer to \cite{Gu:2021xaj}. The scramblon mode propagator $\lambda$ is expected to take the universal form for generic lyapunov exponent $\varkappa$: 
\be
\lambda = -\frac{e^{i\frac{\varkappa}{2}\left(\pi+\theta_3+\theta_4-\theta_1-\theta_2\right)}}{C}
\ee
where $C$ is a normalization constant proportional to $N$; while the explicit form of the vertex function $\Upsilon_m(\theta_{12})$ is less understood except at the maximal chaos $\varkappa=1$, where it is explicitly given by: 
\bea\label{eq:max_chaos}
\Upsilon^m\left(\theta_{12}\right)&=&\int^\infty_0 dy\; y^m\; h(y,\theta_{12}),\;\;
h(y,\theta_{12})=\frac{G}{\Gamma(2\Delta)}\;y^{2\Delta-1} e^{-\Theta_{12} y},\;\;G= \frac{1}{2}\cos{\left(\frac{\pi v}{2}\right)}^{2\Delta},\;\;\Theta_{12} = \cos{\left[\frac{v(\pi - \theta_{12})}{2}\right]}
\eea
where $\Delta =1/q$ is the conformal dimension of $\chi$ in $q$-body SYK models, and $v$ is given by:
\be 
\frac{\pi v}{\cos{\left(\frac{\pi v}{2}\right)}}=\beta J
\ee
For our purpose we will always focus on the strong coupling limit $v\to 1$. The OTOC can be expressed using these as: 
\be\label{eq:SYK_OTOC_max}
OTOC = \sum^\infty_{m=0} \Upsilon^{m}(\theta_{12}) \frac{(\lambda)^m}{m!} \Upsilon^{m}(\theta_{34}) 
\ee
The usual OTOC corresponds to setting: 
\be
\theta_1 = \frac{2\pi i}{\beta}t+\frac{\pi }{2},\;\; \theta_2=\frac{2\pi i}{\beta}t+\frac{3\pi }{2},\;\;\theta_3=\pi,\;\;\theta_4= 0
\ee
For early time $t\ll \log{N},\;\lambda \propto e^{\varkappa t}/N \ll 1$, and the OTOC is dominated by single scramblon exchange at $m=1$, giving the exponential behavior in time. We are interested in computing the operator size generating function $Z(\mu,t_1,t_2)$ defined by: 
\bea
Z(\mu,t_1,t_2) &=& \left\langle \chi_1(t_1), e^{-\mu \hat{N}} \chi_1(t_2)\right\rangle_{TFD} \nonumber\\ 
&=& e^{-\frac{\mu N}{2}}\Big\langle I\Big|e^{-\frac{\beta}{8}\left(\hat{H}_L+\hat{H}_R\right)}\chi^L_1(t_1) e^{-\frac{\beta}{4}\hat{H}_L} \sum_n \frac{1}{n!}\left(-\frac{i\mu}{2} \sum_i \chi^L_i \chi^R_i\right)^n \nonumber\\
&\times& e^{-\frac{\beta}{4}\hat{H}_L} \chi^L_1(t_2) e^{-\frac{\beta}{8}\left(\hat{H}_L+\hat{H}_R\right)}\Big|I\Big\rangle 
\eea
We can again turn this expansion into correlation functions in terms of $\chi^L=\chi$ only. To this end, we need to shift $\chi^R_i$ in the size operator all the way to the left and use the identity (\ref{eq:I_identity}) to transform it into $\chi^L_i$. The key steps proceed as follows:
\bea
&&\left(\sum_i \chi^L_i \chi^R_i\right)^n e^{-\frac{\beta}{4}\hat{H}_L} \chi^L_1(t_2) e^{-\frac{\beta}{8}\left(\hat{H}_L+\hat{H}_R\right)}\Big|I\Big\rangle \nonumber\\
&=& (-1)^{\frac{n(n-1)}{2}}\sum_{i_1,i_2,...,i_n}\left(\chi^L_{i_1}...\chi^L_{i_n}\right)\left(\chi^R_{i_n}...\chi^R_{i_1}\right) e^{-\frac{\beta}{4}\hat{H}_L} \chi^L_1(t_2) e^{-\frac{\beta}{8}\left(\hat{H}_L+\hat{H}_R\right)}\Big|I\Big\rangle \nonumber\\
&=& (-1)^{\frac{n(n+1)}{2}}\sum_{i_1,i_2,...,i_n}\left(\chi^L_{i_1}...\chi^L_{i_n}\right) e^{-\frac{\beta}{4}\hat{H}_L} \chi^L_1(t_2) e^{-\frac{\beta}{8}\hat{H}_L} \left(\chi^R_{i_n}...\chi^R_{i_1}\right)e^{-\frac{\beta}{8}\hat{H}_R}\Big|I\Big\rangle \nonumber\\
&=& (-i)^{n}\sum_{i_1,i_2,...,i_n}\left(\chi^L_{i_1}...\chi^L_{i_n}\right) e^{-\frac{\beta}{4}\hat{H}_L} \chi^L_1(t_2) e^{-\frac{\beta}{4}\hat{H}_L} \left(\chi^L_{i_1}...\chi^L_{i_n}\right)\Big|I\Big\rangle 
\eea
where in the second last line we have used $\left(\hat{H}_R-\hat{H}_L\right) |I\rangle = 0$. Assemble everything, in the end we get the following expression:  \bea
Z(\mu,t_1,t_2)&=& e^{-\frac{\mu N}{2}}\sum_{n}\frac{1}{n!}\Big\langle \mathcal{T}\chi_1\left(t_1+\frac{3i\beta}{4}\right) \left[-\frac{\mu}{2} \sum_i\chi_i\left(\frac{i\beta}{2}\right)\chi_i(0)\right]^n\nonumber\\
&\times& \chi_1\left(t_2+\frac{i\beta}{4}\right) \Big\rangle_\beta
\eea 
where $\langle ... \rangle=\text{Tr}\left(\rho...\right)$ is the thermal correlator and $\mathcal{T}$ denotes time-ordering in the imaginary time. This is a generalization of the OTOC. Through the effective model, this can be expressed in terms of $\Upsilon^{R/A,m}$ and $\lambda$: 
\bea\label{eq:SYK_generating}
Z(t_1,t_2,\mu) &=e^{-\mu N \left(\frac{1}{2}-G(\theta_{34})\right)}\sum^{\infty}_{m=0} \Upsilon^m\left(\theta_{12}\right)\left(\frac{\lambda\mu N}{2}\right)^m\frac{1}{m!} \Upsilon^1\left(\theta_{34}\right)^m
\eea
It is worth making a few comments regarding (\ref{eq:SYK_generating}): (i) we have included in the prefactor a factor of $e^{\mu N G(\theta_{34})}$ to factor out the disconnected contribution to the size operator $\hat{N}$ in the generating function, where the Green's function is given as: 
\be
G\left(\theta_{12}\right)= \frac{1}{2}\left[\frac{\cos{\left(\frac{\pi v}{2}\right)}}{\Theta_{12}}\right]^{2\Delta}
\ee
(ii) the vertex function insertions $\Upsilon^1(\theta_{34})$ related to the size operator only contains those associated with single scramblon emissions $m=1$, this is the leading order contribution in the time regime $t\ll \log{N}$; (iii) the complexified time insertions are given explicitly by: 
\bea\label{eq:imaginary_spacing}
\theta_1 = \frac{2\pi i}{\beta}t_1+\frac{\pi}{2},\;\;\theta_2 =  \frac{2\pi i}{\beta}t_2+\frac{3\pi}{2},\;\;\theta_3=0,\;\;\theta_4 = \pi
\eea
To compute $Z(t_1,t_2,\mu)$ explicitly using (\ref{eq:SYK_generating}) for generic $\varkappa$, we need the modification of the vertex function away from that of $\varkappa =1$ given in (\ref{eq:max_chaos}). We can deduce this by examining how the structure of the OTOC is modified. At maximal chaos $\varkappa$ it can be written by plugging (\ref{eq:max_chaos}) into (\ref{eq:SYK_OTOC_max}) as: 
\bea
OTOC &=& \int^\infty_0 dy_1 \int^\infty_0 dy_2 h\left(y_1,\theta_{12}\right)h\left(y_2,\theta_{34}\right)e^{-\lambda y_1 y_2}\nonumber\\
&=& \int^\infty_0 dy_1 \int^\infty_0 dy_2 h\left(y_1,\theta_{12}\right)h\left(y_2,\theta_{34}\right)e^{\frac{e^t}{C} y_1 y_2}
\eea
where in the second line we have assigned $t_1=t_2=t,\;t_3=t_4=0$ according to the usual OTOC conventions. The modifications to the OTOCs at submaximal chaos $\varkappa<1$ has been studied, e.g. in contexts such as including the stringy corrections\cite{Maldacena:2015waa,Shenker:2014cwa}. The following form of modification was proposed:
\be 
\widetilde{OTOC} = \int^\infty_0 dy_1 \int^\infty_0 dy_2\; h\left(y_1,\theta_{12}\right)h\left(y_2,\theta_{34}\right)e^{\frac{e^{\varkappa t}}{C} (y_1 y_2)^\varkappa} 
\ee 
We shall assume that the form of modification can be extended to generic time insertions $(t_1,t_2,t_3,t_4)$. Then we can obsorb the modifications by re-writing: 
\bea\label{eq:submax_chaos}
\widetilde{OTOC} &=& \int^\infty_0 dy_1 \int^\infty_0 dy_2\; \tilde{h}\left(y_1,\theta_{12}\right)\tilde{h}\left(y_2,\theta_{34}\right)e^{-\tilde{\lambda} y_1 y_2} \nonumber\\
\tilde{\lambda} &=&   -\frac{e^{i\frac{\varkappa}{2}\left(\pi+\theta_3+\theta_4-\theta_1-\theta_2\right)}}{C},\;\;\tilde{h}\left(y,\theta_{ij}\right) = \frac{y^{1/\varkappa-1}}{\varkappa} h\left(y^{1/\varkappa},\theta_{ij}\right)
\eea
In other words, at submaximal chaos $\varkappa<1$ we can work with the modified scramblon mode propagator $\tilde{\lambda}$, as well as the vertex function derived from the modified kernel $\tilde{h}$ in (\ref{eq:submax_chaos}). In what follows, we shall use (\ref{eq:submax_chaos}) to proceed with the computations for generic $\varkappa$. In terms of these, the generating function $Z(t_1,t_2,\mu)$ can be written as: 
\bea
Z(t_1,t_2,\mu)=e^{-\mu N \left(\frac{1}{2}-G(\theta_{34})\right)}\int^\infty_0 dy\;\tilde{h}\left(y,\theta_{12}\right)\exp\left(\frac{\tilde{\lambda} \mu N \Upsilon^1(\theta_{34})}{2}y\right)
\eea
For the insertions (\ref{eq:imaginary_spacing}), the ingredients are given explicitly as follows: 
\bea
\Theta_{12} &=& \cosh{\left(\frac{\pi t_{12}}{\beta}\right)},\;\;t_{12}=t_1 - t_2,\;\;\Theta_{34} = 1 \nonumber\\
\lambda &=& -\frac{1}{C}e^{\frac{\pi\varkappa}{\beta}(t_1+t_2)},\;\;\Upsilon^1(\theta_{34})=\frac{\Gamma(2\Delta+\varkappa)}{\Gamma(2\Delta)} G,\;\;G\left(\theta_{34}\right)= G
\eea 
Plugging these in, we obtain the following explicit form: 
\bea
Z(t_1,t_2,\mu) &=& \frac{e^{-\mu N\left(\frac{1}{2}-G\right)}G}{\Gamma(2\Delta)}\int^\infty_0 \frac{dy}{\varkappa}\; y^{\frac{2\Delta}{\varkappa}-1}\exp\left[-\mu K e^{\frac{\varkappa\pi(t_1+t_2)}{\beta}}y-\cosh{\left(\frac{\pi t_{12}}{\beta}\right)}y^{1/\varkappa}\right],\;\;K=\frac{\Gamma(2\Delta+\varkappa) \mu N G}{2\Gamma(2\Delta)C}
\eea
We could now apply an inverse laplace transform and obtain: 
\bea
&&P(t_1,t_2,\ell) = \frac{1}{2\pi i}\oint_{\Gamma}d\mu\; e^{\mu \ell}  Z(\mu,t_1,t_2)\nonumber\\
&=& \frac{G}{\Gamma(2\Delta)}\int^\infty_0 \frac{dy}{\varkappa}\; \delta\left(\tilde{\ell}-Ke^{\frac{\pi \varkappa}{\beta}(t_1+t_2)}y\right)\;y^{\frac{2\Delta}{\varkappa}-1}e^{-\cosh{\left(\frac{\pi t_{12}}{\beta}\right)}y^{1/\varkappa}}\nonumber\\
&=& \frac{G\;\tilde{\ell}^{2\Delta/\varkappa-1}}{\varkappa K^{2\Delta/\varkappa}\Gamma(2\Delta)}\exp\left[-\frac{2\pi\Delta}{\beta}\left(t_1+t_2\right)-\frac{1}{2}\left(\frac{\tilde{\ell}}{K}\right)^{1/\varkappa}\left(e^{-\frac{2\pi t_1}{\beta}}+e^{\frac{-2\pi t_2}{\beta}}\right)\right]
\eea
where we have defined the renormalized operator-size $\tilde{\ell}$:
\be
\tilde{\ell}=\ell-N\left(\frac{1}{2}-G\right)
\ee
We make some observations. Firstly the inverse laplace transform $P(t_1,t_2,\ell)$ depends on $\ell$ only through the combination $\left(\frac{\tilde{\ell}}{K}\right)^{1/\varkappa}$, which can be viewed as an effective operator-size for $\varkappa <1$. Secondly, the dependence on the two time insertions $\lbrace t_1,t_2\rbrace$ of $P(t_1,t_2,\ell)$ can be factorized:
\be 
P(t_1,t_2,\ell) = Q_{\ell}(t_1)\times Q_{\ell}(t_2),\;\;\; Q_{\ell}(t) = \frac{\tilde{\ell}^{\Delta/\varkappa-1/2}}{\sqrt{\varkappa} K^{\Delta/\varkappa}}\sqrt{\frac{G}{\Gamma(2\Delta)}}e^{-\frac{2\pi\Delta}{\beta}t-\frac{1}{2}\left(\frac{\tilde{\ell}}{K}\right)^{1/\varkappa}e^{-\frac{2\pi}{\beta}t}}
\ee
As a consequence, $K_{mn}(\ell)$ also factorizes: 
\be
K_{mn}(\ell) =\ell J_m(\ell) J_n(\ell)  
\ee
It is interesting to contemplate what is behind the observed factorization property of $K_{mn}(\ell)$. To see what is happening, let us write it as a product of rectangular matrices:
\be
\ell^{-1}K_{mn}(\ell) = \left[L(\ell)^\dagger\cdot L(\ell)\right]_{mn} = J_m(\ell) J_n(\ell), \;\;\; \left[L(\ell)\right]_{m\alpha} = \langle O_m, \alpha \rangle,\;\;\alpha \in H_{\ell} 
\ee
where $H_{\ell}$ denotes the operator space sector with fixed size $\ell$.  Factorization thus implies that the rectangular matrix $L(\ell)$ is of rank one. We can therefore deduce that:
\be
 \langle O_m, \alpha \rangle = \left[L(\ell)\right]_{m\alpha} = J_m(\ell) \times \Psi^\ell_\alpha \rightarrow\; \hat{P}(\ell)|O_m\rangle = J_m(\ell) \times |\Psi^\ell\rangle,\;\;|\Psi^\ell\rangle =\sum_{\alpha \in \mathcal{H}_\ell} \Psi^\ell_\alpha |\alpha\rangle
\ee 
In other words, distinct Krylov basis states $\mathcal{O}_{m}$ share the same normalized projection $|\Psi^\ell\rangle$  into each operator space sector $ H_\ell$.  This is natural from the perspective that the Hamiltonian is symmetric under the permutation $P(n)$, the action of which is ``ergodic" in each sector $H_\ell$.  We therefore expect the common projection $|\Psi^\ell\rangle$ to be the most permutation-symmetric operator state in $H_\ell$. The factors $J_m(\ell)$ can therefore be viewed as the wave-function in operator-size of $\mathcal{O}_m$. 

\begin{figure}[t]
\centering
\includegraphics[width=0.92\linewidth]{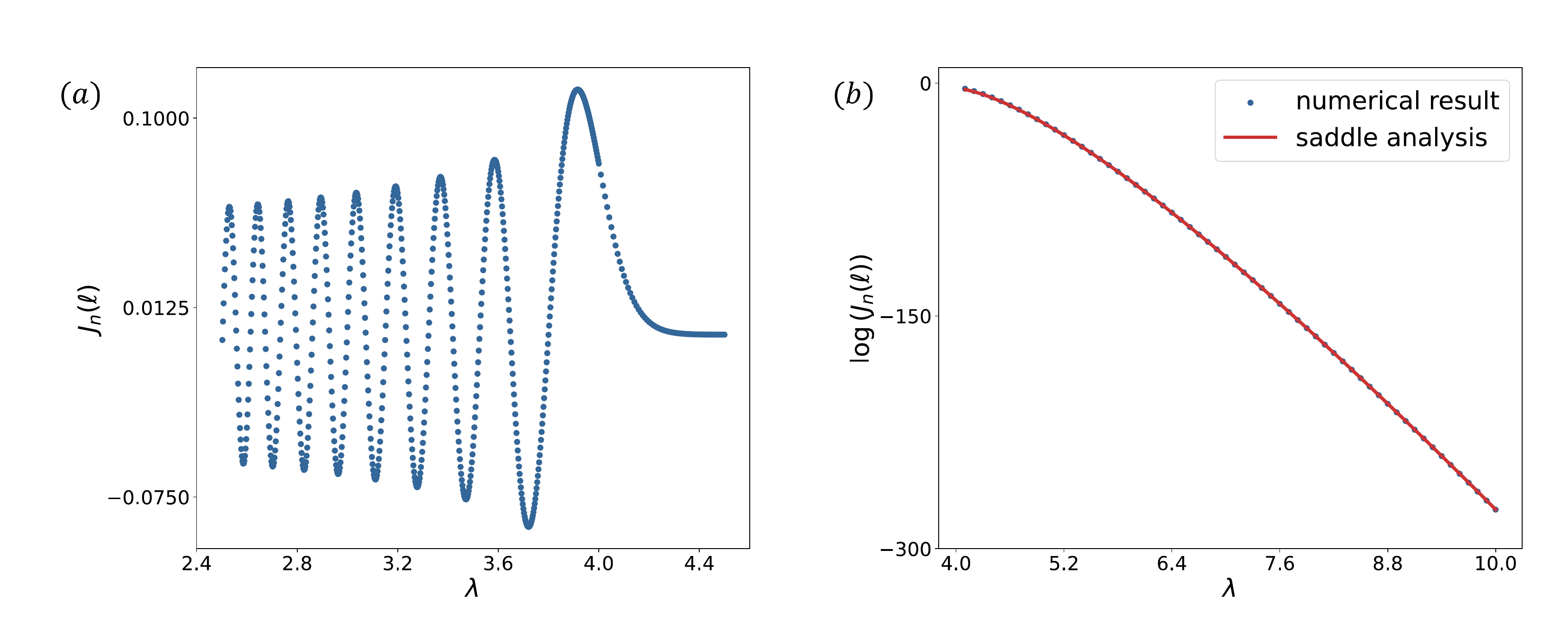}
\caption{ We present the  asymptotic behaviors of $J_{n}(\ell)$ as functions of $\lambda=\frac{\left(\ell/K\right)^{1/\varkappa}}{n}$ for $n=150$:  (a). the numerical result showing a transition between oscillatory and exponential decay across $\lambda =4$;  (b). the numerical result vs. saddle analysis in the exponential decay regime $\lambda > 4$. }
\label{fig:AsymJn}
\end{figure}

In the case of $\varkappa=1$, we can therefore compute $J_m(\ell)$ from $Q_\ell(t)$ via a single contour integral. Going through some algebra produces the following explicit integral:
\be\label{eq:size_krylov_1}
J_n(\ell) = \frac{\tilde{\ell}^{\Delta/\varkappa-1/2}}{\sqrt{\varkappa}K^{\Delta/\varkappa}}\sqrt{\frac{\Gamma(n+1)G}{\Gamma(2\Delta+n)}} \oint \frac{dy}{y^{n+1}}\left(1+y\right)^{-2\Delta}  e^{-\frac{1}{2}\left(\frac{\tilde{\ell}}{K}\right)^{1/\varkappa}\left(\frac{1-y}{1+y}\right)}
\ee
To extract the global behavior of $J_n(\ell)$ at large $n$, it is appropriate to rescale $\left(\ell/K\right)^{1/\varkappa}=\lambda n$. The saddle-point equation and the solutions are then given by: 
\bea
     n  + \frac{2\Delta  y}{1+y} -\frac{\lambda n y}{(1+y)^{2}} = 0,\;\; y^{*}_{\pm}=\frac{n\left(\lambda-2\right)- 2 \Delta \pm \sqrt{(n\lambda-2 \Delta)^2-4 n^2\lambda}}{2(n+2 \Delta)}
\eea
It turns out that the physical saddle corresponds to $y^*_-$, while the other saddle $y^*_+$ gives a wave-function $J_n(\ell)$ that grows exponentially with $\ell$. Now plugging this in and taking into account the scaling factor from the fluctuations, we obtain the following asymptotic behavior of $J_{n}(\tilde{\ell} )$:
 \begin{equation}\label{eq:SYK_size1}
     \begin{aligned}
         J_{n}(\tilde{\ell}) \sim  \frac{\tilde{\ell}^{\Delta-1/2}}{K^{\Delta}} \sqrt{\frac{\Gamma(n+1)}{\Gamma(2\Delta+n)}}  e^{- S^{*}} \times (\text{fluctuation})  \sim   n^{-1/2 }  \left[ e^{\frac{ \sqrt{\lambda (\lambda-4)}}{2}} \left(\frac{\lambda -2  - \sqrt{\lambda (\lambda -4)}}{2}\right) \right]^{-n}
     \end{aligned}
 \end{equation} 
This result is consistent with the numerical calculation for $\lambda > 4$, see FIG. \ref{fig:AsymJn} (b). From this, we see that the wave-function $J_n(\ell)$ exhibits a transition across $\lambda \sim 4$: in terms of the original renormalized operator size $\tilde{\ell}$,  the wave-function is oscillatory in the regime: $\tilde{\ell}< K\left(4n\right)^\varkappa$ ; and decays exponentially in the regime: $\tilde{\ell}>K\left(4n\right)^{\varkappa}$, see FIG. \ref{fig:AsymJn} (a). As a result, one can estimate the typical operator size of $\mathcal{O}_n$ to be of order: $\tilde{\ell}\sim K n^{\varkappa}$.

 \subsection{Example 2: MBL systems}
 Next, let us consider the operator-size distribution of MBL systems. We begin with the Heisenberg evolution of $\sigma^0_x$. Applying the identity (\ref{eq:pauli_idenity_1}) introduced previously, we can derive: 
\begin{equation}
    \sigma_x^0(t)=e^{iHt}\sigma^0_x e^{-iHt}=\sigma_x^0\prod_{j\neq 0} \left(\cos{(2J_{0j}t)}+i\sin{(2J_{0j}t)\sigma^0_z \sigma^{j}_z}\right)\left(\cos(2h_0 t)+i\sin(2h_0 t)\sigma_z^0\right).
\end{equation}
The generating function $Z(t_1,t_2,\mu)$ now is defined to be: 
\bea
Z(t_1,t_2,\mu) &=&\text{Tr}\left[\sigma^0_x(t_1) e^{-\mu \hat{N}}\sigma^0_x(t_1)\right]\nonumber\\
&=&\text{Tr} \Bigg\lbrace\sigma_x^0\prod_{j\neq 0} \left(\cos{(2J_{0j}t_1)}+i\sin{(2J_{0j}t_1)\sigma^0_z \sigma^{j}_z}\right)\left(\cos(2h_0 t_1)+i\sin(2h_0 t_1)\sigma_z^0\right)\nonumber\\
&\times & e^{-\mu \hat{N}}\sigma_x^0\prod_{k\neq 0} \left(\cos{(2J_{0k}t_2)}+i\sin{(2J_{0k}t_2)\sigma^0_z \sigma^{k}_z}\right)\left(\cos(2h_0 t_2)+i\sin(2h_0 t_2)\sigma_z^0\right)\Bigg\rbrace
\eea
This can be computed by direct counting techniques. Let us study the pattern of Pauli strings that arise from taking products within either pf $\sigma^0_x(t_{1,2})$. The following observations arise: (i) each factor of $\cos{(2J_{0j}t)}$ corresponds to an identity operator on site $j$, while each factor of $\sin{(2J_{0j}t)}$ indicates the existence of a $\sigma_z$ operator on site $j$; (ii) terms containing $\cos(2h_0 t)$ or $\sin(2h_0 t)$ differ by interchanging $\sigma_x^0$ and $\sigma_y^0$. Taking trace then pairs up Pauli strings from  $\sigma^0_x(t_{1,2})$. Therefore, We can neglect all cross terms between sine and cosine functions when computing the generating function, which gives:
\begin{equation}
    \begin{aligned}
        Z(\mu,t_1,t_2)=&e^{-\mu}\cos\left(2h_0 (t_1-t_2)\right)\prod_{j\neq 0}\left(\cos(2J_{0j} t_1)\cos(2J_{0j} t_2)+e^{-\mu}\sin{(2J_{0j}t_1)}\sin{(2J_{0j}t_2)}\right).
    \end{aligned}
\end{equation} 
The first term $e^{-\mu}$ comes from the zeroth site, which is always occupied. Additional factor of $e^{-\mu}$ appears when other sites are occupied by $\sigma_z$. After taking the disorder average, it becomes
\begin{equation}
    \begin{aligned}
          \overline{ Z(\mu,t_1,t_2) }
          &= e^{-\mu} \overline{\cos(2h_0 (t_1-t_2)) }\prod_{j\neq 0} \bigg[\frac{1-e^{-\mu}}{2}\overline{\cos{\left(2_{0j}(t_1+t_2)\right)}}+\frac{1+e^{-\mu}}{2}\overline{\cos{\left(2_{0j}(t_1-t_2)\right)}} \bigg] \\
          &=e^{-\mu}e^{-2h^2(t_1-t_2)^2}\prod_{j\neq 0}\left(\frac{1-e^{-\mu}}{2}e^{-2J^2(t_1+t_2)^2e^{-\frac{|j|}{\xi}}}+\frac{1+e^{-\mu}}{2}e^{-2J^2(t_1-t_2)^2e^{-\frac{|j|}{\xi}}}\right). 
    \end{aligned}
\end{equation}
Aiming at extracting $K_{mn}$, we multiply the generating function by $e^{\gamma^2(t_1^2+t_2^2)/2}$, with $\gamma^{2} = 4 J^{2} \sum_{j\neq 0} e^{-\frac{|j|}{\xi} } + 4h^{2}$ introduced previously. The result can be simplified as 
\begin{equation}
   \overline{ Z(\mu,t_1,t_2) } e^{\frac{\gamma^2}{2}(t_1^2+t_2^2)}=e^{\gamma^{2} t_1t_2} e^{-\mu}\prod_{j\neq 0}\left(\frac{1+e^{-\mu}}{2}+\frac{1-e^{-\mu}}{2}e^{-8J^2t_1t_2e^{-\frac{|j|}{\xi}}}\right).
\end{equation}
Again, since this is only a function of $t_1 t_2$, we expect $K_{mn}(\ell)$ to be exactly diagonal for all $\ell$. We can approximate the product via the similar logic  used previously: when $8J^2t_1t_2e^{-\frac{|j|}{\xi}}\gg 1$, the bracket gives $(1+e^{-\mu})/2$, indicating an equal probability between $\sigma^j_z$ and $I$; for $8J^2t_1t_2e^{-\frac{|j|}{\xi}}\ll 1$, the bracket is $1$, which indicates a trivial identity operator. Under this approximation, we have that:
\begin{equation}
    \overline{ Z(\mu,t_1,t_2) } e^{\frac{\gamma^2}{2}(t_1^2+t_2^2)} \approx e^{\gamma^{2} t_1t_2} e^{-\mu} \left(\frac{1+e^{-\mu}}{2}\right)^{2 \xi \ln (8J^2t_1t_2) } = e^{\gamma^2 t_1 t_2}2^{-M(t_1 t_2)} \sum_{\ell}B\left(M(t_1t_2),\ell\right)e^{-\mu(\ell+1)} = \sum_{\ell}e^{-\mu \ell}\ell^{-1}\sum_n \frac{(\gamma^2 t_1t_2)^n}{n!}  K_{nn}(\ell)
\end{equation}
where $M(t_1 t_2)= 2 \xi \ln (8J^2t_1t_2)$ and $B\left(M,\ell\right)$ is the binomial coefficient. The resolution $K_nn(\ell)$ of the  diagonal Krylov metric into fixed operator-size $\ell$ can therefore be written in $x=\gamma^2 t_1 t_2$ as:
\begin{equation}
  K_{nn}(\ell) \approx \frac{n! \ell}{2\pi i} \oint \frac{dx}{x^{n+1}} \left[\frac{B\left(M(x/\gamma^2),\ell-1\right)}{2^{M(x/\gamma^2)}}\right]e^{x}
\end{equation} 
For $M,\ell$ both large and of the same order, the binomial coefficients approaches a Gaussian distribution:
\be
\frac{B\left(M,\ell\right)}{2^M} \approx \left(\pi M/2\right)^{-1/2} e^{-\frac{\left(\ell-M/2\right)^2}{M/2}} 
\ee
Giving the following integral expression that allows the saddle-point approximation: 
\be
K(\ell) \approx \frac{n!\ell}{2\pi i} \oint dx\; \exp\left(-(n+1)\ln{x}+x- \frac{\left(\ell-\xi \ln{\left(\frac{8J^2}{\gamma^2} x\right)}\right)^2}{\xi\ln{\left(\frac{8J^2}{\gamma^2}x\right)}}-\frac{1}{2}\ln{\left(\pi \xi\ln{\left(\frac{8J^2}{\gamma^2} x\right)}\right)}\right)
\ee 
To access the global distribution, we can rescale $\ell = \lambda \ln{n}$. The dominant saddle-point solutions in this limit is simply: 
\be
x^* = n + ..
\ee
Plugging this back into the integral and taking care of the additional factor from the fluctuations, we obtain in the end the asymptotic form: 
\be
K_{nn}(\ell) \sim \lambda \exp\left(-\frac{(\lambda-\xi)^2}{\xi}\ln{n}\right)
\ee
We end with a few comments. The operator-size distribution of $\mathcal{O}_n$ is given by a Gaussian with the average size: 
\be
\overline{\ell} \sim \xi \ln{n}
\ee
It is also worth understanding the diagonal property of $K_{mn}(\ell)$ for any $\ell$. In particular, it implies that: 
\be 
K_{mn}(\ell) = \ell\langle \mathcal{O}_m |\;\hat{P}(\ell)\;|\mathcal{O}_n\rangle = \ell\langle \hat{P}(\ell) \mathcal{O}_m,\hat{P}(\ell)\mathcal{O}_n\rangle   \propto \delta_{mn} 
\ee
In other words, distinct Krylov basis elements project onto mutually orthogonal states in each sector $H_{\ell}$. This reflects the localized nature of the dynamics as it explores within each sector of the operator space with fixed size $\ell$.

\end{document}